\documentclass[
reprint,            
superscriptaddress, 
amsmath,            
amssymb,            
aps,                
prd,                 
floatfix,           
nofootinbib,        
]{revtex4-1}

\usepackage{tensor}     
\usepackage{graphicx}   
\usepackage[
colorlinks=true,        
citecolor=blue,         
linkcolor=blue,         
urlcolor=blue           
]{hyperref}             
\usepackage{bm}         
\usepackage{xcolor}     
\newcommand{\newc}{\newcommand*}
\newc{\figurewidth}{3.2in}
\newc{\xbar}{\bar{x}}
\newc{\rhoeq}{\rho_{\rm{eq}}}
\newc{\zeq}{z_{\rm{eq}}}
\newc{\la}{\lambda}
\newc{\tla}{\tilde{\la}}
\newc{\dt}{\delta}
\newc{\Dt}{\Delta}
\newc{\vj}{\vec{j}}
\newc{\vl}{\vec{l}}
\newc{\hx}{\hat{x}}
\newc{\hy}{\hat{y}}
\newc{\bj}{\bm{j}}
\newc{\mJ}{\mathcal{J}}
\newc{\mP}{\mathcal{P}}
\newc{\ga}{\gamma}
\newc{\Msun}{M_\odot}
\newc{\app}{\approx}
\newc{\av}[1]{\langle #1 \rangle}
\newc{\eq}[1]{Eq.~\eqref{#1}}
\newc{\al}{\alpha}
\newc{\Xstar}{X_{\ast}}
\newc{\seq}{\sigma_{\rm{eq}}}
\newc{\fpbh}{f_{\rm{pbh}}}
\newc{\RR}{{\cal R}}
\newc{\lvc}{LIGO/Virgo}
\newcommand{\nc}{\newcommand*}

\nc{\Eq}[1]{Eq.~\eqref{#1}}     
\nc{\Fig}[1]{Fig.~\ref{#1}}     
\nc{\Table}[1]{Table~\ref{#1}}  
\nc{\Sec}[1]{Sec.~\ref{#1}}     

\def\({\left(}
\def\){\right)}
\def\[{\left[}
\def\]{\right]}

\def\e{\begin{equation}}
\def\q{\end{equation}}
\def\m{\begin{eqnarray}}
\def\n{\end{eqnarray}}

\usepackage{color}

\begin{document}
\title{Gravitational and electromagnetic radiation from binary black holes with electric and magnetic charges: Circular orbits on a cone}
\author{Lang Liu}
\email{liulang@itp.ac.cn}
\affiliation{CAS Key Laboratory of Theoretical Physics,
Institute of Theoretical Physics, Chinese Academy of Sciences,
Beijing 100190, China}
\affiliation{School of Physical Sciences,
University of Chinese Academy of Sciences,
No. 19A Yuquan Road, Beijing 100049, China}

\author{{\O}yvind Christiansen}
\email{oyvind.christiansen@astro.uio.no}
\affiliation{Institute of Theoretical Astrophysics, University of Oslo, Sem S$\ae$lands vei 13,0371 Oslo, Norway}
\author{Zong-Kuan Guo}
\email{guozk@itp.ac.cn}
\affiliation{CAS Key Laboratory of Theoretical Physics,
Institute of Theoretical Physics, Chinese Academy of Sciences,
Beijing 100190, China}
\affiliation{School of Physical Sciences,
University of Chinese Academy of Sciences,
No. 19A Yuquan Road, Beijing 100049, China}
\affiliation{School of Fundamental Physics and Mathematical Sciences,
Hangzhou Institute for Advanced Study,
University of Chinese Academy of Sciences, Hangzhou 310024, China}
\author{Rong-Gen Cai}
\email{cairg@itp.ac.cn}
\affiliation{CAS Key Laboratory of Theoretical Physics,
Institute of Theoretical Physics, Chinese Academy of Sciences,
Beijing 100190, China}
\affiliation{School of Physical Sciences,
University of Chinese Academy of Sciences,
No. 19A Yuquan Road, Beijing 100049, China}
\affiliation{School of Fundamental Physics and Mathematical Sciences,
Hangzhou Institute for Advanced Study,
University of Chinese Academy of Sciences, Hangzhou 310024, China}
\author{Sang Pyo Kim}
\email{sangkim@kunsan.ac.kr}
\affiliation{CAS Key Laboratory of Theoretical Physics,
Institute of Theoretical Physics, Chinese Academy of Sciences,
Beijing 100190, China}
\affiliation{Department of Physics, Kunsan National University, Kunsan 54150, Korea}
\date{\today}

\begin{abstract}
 We derive the equations of motion of dyonic black hole binaries with electric and magnetic charges and explore features of static orbits. By using a Newtonian method with the inclusion of radiation reaction, we calculate the total emission rate of energy and angular momentum due to gravitational radiation and electromagnetic radiation for circular orbits. Moreover, we obtain the evolutions of orbits and calculate merger times of dyonic binaries. We  show that electric and magnetic charges significantly suppress the merger times of dyonic binaries.  Our results provide rich information about dyonic binaries and can be used to test black holes with magnetic charges.
\end{abstract}

\maketitle

\section{Introduction}\label{intro}

The first direct detection of gravitational waves (GWs) from a binary  black hole
coalescence \cite{Abbott:2016blz} has opened a new window of physics and astronomy.
Ten merger events of binary black holes have been reported by \lvc\ during
the O1 and O2 observing runs over the past few years \cite{Abbott:2016blz,Abbott:2016nmj,
TheLIGOScientific:2016pea,Abbott:2017vtc,
Abbott:2017gyy,Abbott:2017oio,LIGOScientific:2018mvr}.
The progenitors of these binaries are under intensive investigation  but still unknown \cite{Bird:2016dcv,Sasaki:2016jop,
Clesse:2016vqa,Ali-Haimoud:2017rtz,Belczynski:2016obo}.  These \lvc\ black holes show a much heavier mass distribution than the mass distribution
inferred from X-ray observations
\cite{Casares:2013tpa,Corral-Santana:2013uua,%
Corral-Santana:2015fud},
which presents a gigantic challenge to the formation and evolution mechanisms of astrophysical
black holes. One possible explanation for \lvc\ black holes is that they are primordial black holes (PBHs)
\cite{Bird:2016dcv,Sasaki:2016jop,Ali-Haimoud:2017rtz,Chen:2018czv,Raidal:2018bbj,Liu:2018ess,Liu:2019rnx}
formed in the radiation-dominated era of the early universe due to the collapse of large energy density fluctuations \cite{Hawking:1971ei, Carr:1974nx}. Besides being the sources of \lvc~detection, PBHs can also be a candidate for dark matter or the seeds for galaxy formation \cite{Bean:2002kx,Kawasaki:2012kn,Carr:2018rid}.

Black holes can have additional hairs of electric and magnetic charges besides  the mass and angular momentum. These charged black holes have rich phenomena  compared to uncharged black holes. In recent years, charged black holes have attracted a lot of attention and have been studied extensively \cite{Cardoso:2016olt,Liebling:2016orx,Bai:2019zcd,Liu:2020cds,Christiansen:2020pnv,Wang:2020fra,Bozzola:2020mjx}. For  GW150914, the first merger event of binary black holes reported by \lvc,  it is shown in Refs.~\cite{Wang:2020fra,Bozzola:2020mjx} that these black holes can have some electric charge, while Ref.~\cite{Liebling:2016orx} argues that the magnetic charged black holes  have to be small. A binary of black holes with charges emits not only GWs but also electromagnetic waves. For a binary of black holes with electric charges, the Coulomb force gives an additional central force to the gravitational force and thus modifies the Keplerian orbit by the relative ratio of the Coulomb force to the gravitational force. This ratio affects the power distribution of GWs and leads to a different power-law of the merger rate \cite{Liu:2020cds}. Compared to electric black holes, magnetic black holes can be relatively long-lived and  a possible  candidate for dark matter.  Very recently, Maldacena has argued that the extremely strong magnetic field near the event horizon of a magnetic black hole may restore the electroweak symmetry and affect the phase-transition scenario in the early universe \cite{Maldacena:2020skw}. Magnetic black holes, which can have electroweak-symmetric coronas outside of the event horizon, are studied as PBHs \cite{Bai:2020spd}.

 While purely electric (or magnetic) binary black holes have been studied comprehensively, few works have focused on gravitational and electromagnetic radiation from binary black holes with electric and magnetic charges. In this paper, we investigate binaries of black holes with electric and magnetic charges in the Einstein-Maxwell theory.~A nonrotating black hole with an electric charge $q$ and a magnetic charge $g$, the so-called dyonic black hole, has the same metric as the Reissner-Nordstr\"{o}m black hole with $q^2$ replaced by $q^2+g^2$, and a rotating dyonic black hole has a similar structure \cite{1982PhRvD..25..995K}. A binary of dyonic black holes has a generalized angular momentum, which is a conserved Laplace-Runge-Lenz vector \cite{2004math.ph...3028L}, around which the orbital plane precesses, and which confines the orbit to a cone \cite{poincare1896remarques}.  The binary follows the Keplerian orbits with the conserved energy and the square of angular momentum that are modified by magnetic charges. The emission power of electromagnetic waves and GWs thus drastically changes, which may provide a new window to identify primordial magnetic charges.

 The primary aim of this paper is to investigate gravitational and electromagnetic radiations from binary black holes with electric and magnetic charges in circular orbits. Compared to purely magnetic (or electric) black holes, we show that dyonic black hole inspirals have richer features. For dyonic binaries, in the 0th order post-Newtonian (PN) expansion, a  non-central and angular-momentum-dependent force causes the orbits to display complex and three-dimensional trajectories. We calculate the total emission rate of energy and angular momentum due to gravitational radiation and electromagnetic radiation within the 0th order post-Newtonian (PN) expansion, particularly, in circular orbits (the case of $e=0$). Moreover, we obtain the evolutions of orbits, calculate merger times of dyonic binaries and show that electric and magnetic charges significantly suppress merger times of dyonic binaries. Our methods can be applied to the early inspirals of low-mass binaries in LIGO/Virgo which have a longer signal in the detector-band.  Of course, when black holes approach to the merger stage, a higher-order PN expansion or numerical-relativity simulation is needed. The main region of applicability of our results is the long inspirals that space-based GW detectors, such as LISA~\cite{Audley:2017drz} and Taiji ~\cite{Guo:2018npi}, will detect.

The organization of this paper as follows. In \Sec{EOM}, we derive the equations of motion of dyonic binaries and investigate characteristics of the static orbits. In \Sec{Emission}, we calculate the total emission rate of angular momentum and energy due to gravitational radiation and electromagnetic radiation, derive the evolution of $a$ and $\theta$, and find the merger time. The \Sec{Concl} is devoted to conclusions and discussions. In Appendix, we show the chaotic orbits in general for dynonic binaries.

\section{Solutions without radiation}\label{EOM}

 Using a Newtonian method, we study the orbital motion of a binary of black holes with electric and magnetic charges. Maxwell's equations with magnetic monopoles are given by\footnote{In this paper, we use units of $G=c =4 \pi \varepsilon_{0} = \frac{\mu_0}{4\pi}=1$.}
\m
\left\{\begin{array}{ll}
\boldsymbol{\nabla} \cdot \boldsymbol{E} & =4 \pi \rho_{\mathrm{e}},  \\
\boldsymbol{\nabla} \times \boldsymbol{E} & =-4\pi \boldsymbol{j}_{\mathrm{m}}-\partial \boldsymbol{B} / \partial t ,\\
\boldsymbol{\nabla} \cdot \boldsymbol{B} & =4\pi \rho_{\mathrm{m}}, \\
\boldsymbol{\nabla} \times \boldsymbol{B} & =4\pi \boldsymbol{j}_{\mathrm{e}}+\partial \boldsymbol{E} / \partial t,
\end{array}\right.
\n
where $\rho_m$ is a magnetic charge density and $\bm{j_m}$ is a magnetic current.  The Lorentz force on a dyon with an electric charge $q$ and a magnetic charge $g$ is
\m
\mathbf{F}= q(\mathbf{E}+\mathbf{v} \times \mathbf{B})+
g\left(\mathbf{B}-\mathbf{v} \times \mathbf{E}\right).
\n
A point dyon generates the electric and magnetic fields
\m
\mathbf{E} = q \frac{\mathbf{r}}{r^3}, \quad \mathbf{B} = g \frac{\mathbf{r}}{r^3}.
\n

In this paper we consider a black hole binary with electric and magnetic charges ($q_1$, $g_1$) and ($q_2$, $g_2$). For a nonrotating dyonic black hole, its metric can be 
described by 
\m
\label{dyon met}
ds^2 = -f_i(r) dt^2 + \frac{dr^2}{f_i(r)} + r^2 (d\theta^2+\sin^2 \theta d\phi^2),
\n
where
\m f_i(r) = 1 - \frac{2m_i}{r} + \frac{q_i^2 + g_i^2}{r^2}, \quad (i = 1, 2),
\n
where $m_i$ is the mass of the black hole.

The nonrelativistic interaction of two dyons in the Minkowski spacetime was studied in classical theory~\cite{1976AnPhy.101..451S} and in a quantum theory~\cite{1968PhRv..176.1480Z}. We will study the bounded motion of two dyonic black holes as a binary system. The Keplerian motions with or without an angular momentum-dependent force term are classified \cite{2004math.ph...3028L}.

For a binary of  dynonic black holes, we choose the center of mass system at the origin
\m
r_{1}^{i}=-\frac{m_{2}}{M} R^{i}, \quad r_{2}^{i}=\frac{m_{1}}{M} R^{i},
\n
where
\m
R^{i}=r_{2}^{i}-r_{1}^{i}, \quad M=m_1+m_2.
\n
Considering the Lorentz force and gravitational force, the equation of motion is given by
\m
\label{eom}
m_{2} \ddot{r}_{2}^{i}=\mu \ddot{R}^{i}=C \frac{R^{i}}{R^{3}}-D \epsilon_{ j k}^{i} \frac{R^{j}}{R^{3}} v^{k},
\n
where
\m
\mu=\frac{m_1m_2}{M}, \quad v^{i}=\dot{R}^{i},
\n
and 
\m
C=\left(-\mu M+q_{1} q_{2}+g_{1} g_{2}\right), \  D=\left(q_{2} g_{1}-g_{2} q_{1}\right).
\n
Note that the case of $D=0$, which corresponds to purely electric or magnetic charges or $q_2/q_1= g_2/g_1$  of balancing out the velocity-dependent Lorentz forces, yields the same orbital motion as that of purely electric charges.
Here, we rewrite Eq.~\eqref{eom} as
\m
\mu \bm{\ddot{R}}=C \frac{\bm{R}}{R^{3}}-D \frac{\bm{R}}{R^{3}} \times \bm{v},
\n
and take a cross-product on both sides with $\bm{R}$. Then we get
\m
\mu \epsilon_{j k}^{i} R^{j} \ddot{R}^{k}=\frac{d}{d t}\left(\mu \epsilon_{j k}^{i} R^{j} \dot{R}^{k}\right)=\frac{d}{d t} \tilde{L}^{i},
\n
where $\bm{\tilde{L}}\equiv \mu \bm{R} \times \bm{v}$ is the real angular momentum of the binary system. Writing the right hand side as
\m
\dot{\tilde{L}}^{i}&=&-D \epsilon^{i j k} R_{j} \epsilon_{k l m} \frac{R^{l}}{R^{3}} \dot{R}^{m}
 \notag \\
&=&-\frac{D}{R^{3}}\left(\delta_{l}^{i} \delta_{m}^{j}-\delta_{m}^{i} \delta_{l}^{j}\right) R_{j} R^{l} \dot{R}^{m}
 \notag \\
&=&\frac{D}{R}\left(\dot{R}^{i}-R^{i} \dot{R} / R\right),
\n
and using the relation
\m
\dot{\hat{\bm{r}}}=\frac{d}{d t}\left(\frac{R^{i}}{R}\right) \hat{\bm{x}}_i=\frac{1}{R}\left(\dot{R}^{i}-R^{i} \dot{R} / R\right) \hat{\bm{x}}_i,
\n
where $\hat{\bm{r}}$ is the unit vector along $\bm{R}$ and $\hat{\bm{x}}_i$ is the unit vector along $x_i$ axis, we find a conserved quantity
\m
\label{L1}
\dot{\bm{L}} =0, \quad \bm{L}\equiv \bm{\tilde{L}}-D \hat{\bm{r}}.
\n
Note that $\bm{L}$ is a Laplace-Runge-Lenz vector \cite{2004math.ph...3028L} and has a meaning of the generalized angular momentum of the binary system. From the definition \eqref{L1} and the fact that $\bm{\tilde{L}}$ is perpendicular to $\hat{\bm{r}}$, we can obtain
\m
\label{Relation}
(L)^2=(\tilde{L})^2+D^2.
\n
Since $\bm{L}$ is conserved and $D$ is a constant, we arrive at  a conclusion that the magnitude of the real angular momentum $\tilde{L}$ is conserved even though the direction of $\bm{\tilde{L}}$ changes.

\begin{figure}[htpb]
    \includegraphics[width=0.42\textwidth]{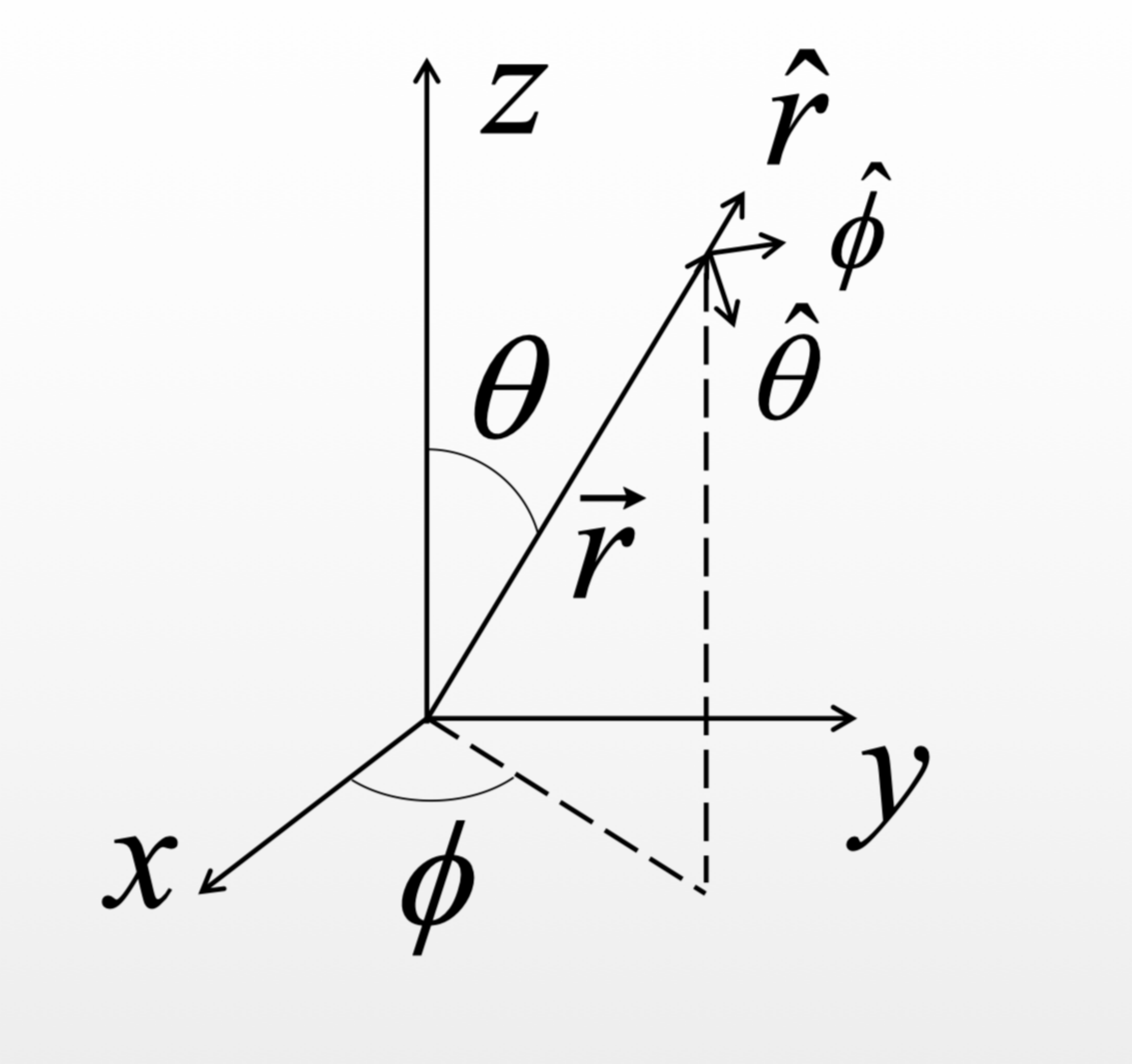}
    \caption[]{Some conventions in a spherical coordinate system.}
    \label{fig:axis}
\end{figure}

Now, we derive the orbit equation of  the binary system.
Because $\bm{L}$ is conserved, we can pick our coordinate system  such that it points along the $z$-axis, that is, $\mathbf{L}$ coincides with the polar axis. Some conventions in a spherical coordinate system are shown in Fig.\ref{fig:axis}. Using $\hat{\mathbf{z}}=\hat{\mathbf{r}} \cos \theta-\hat{\boldsymbol{\theta}} \sin \theta$  in the spherical coordinate $(r, \phi, \theta)$, we have
\m
\boldsymbol{L}=L \hat{\boldsymbol{z}}=L\left(\begin{array}{c}\cos \theta \\ 0 \\ -\sin \theta\end{array}\right)=\tilde{\boldsymbol{L}}-D \hat{\boldsymbol{r}}=\left(\begin{array}{c}
-D \\
\tilde{L}_{\phi} \\
\tilde{L}_{\theta}
\end{array}\right),
\notag \\
\n
and
\m
\cos \theta=-D / L.
\n
Due to both $D$ and $L$ are constants for the binary system, it means that $\theta$ associated with the mass center of the system keeps 
as a constant. 
The special case of $D=0$ gives an orbit on the equatorial plane.
Thus, the orbital equation is given by
\m
\boldsymbol{R}=R\left(\begin{array}{c}
\sin \theta \cos \phi \\
\sin \theta \sin \phi \\
\cos \theta
\end{array}\right).
\n
For the energy of the system, we have
\m
\label{E}
E&=&\frac{1}{2} \mu v^{2}+\frac{C}{R}=\frac{1}{2} \mu(v_{\|}^{2}+v_{\perp}^{2})+\frac{C}{R}
\notag \\
&=&\frac{1}{2} \mu \dot{R}^{2}+\frac{\tilde{L}^{2}}{2 \mu R^{2}}+\frac{C}{R},
\n
where $\bm{v_{\|}}\equiv\dot{R} \hat{\bm{r}}$ is the velocity along $\bm{R}$ and
\m
 \bm{v_{\perp}}\equiv R \dot{\phi} \sin \theta \hat{\boldsymbol{\phi}}+R \dot{\theta} \hat{\boldsymbol{\theta}}=R \dot{\phi} \sin \theta \hat{\boldsymbol{\phi}}
\n
is the velocity that is perpendicular to $\bm{R}$.
To start with, we can solve Eq.~\eqref{E} for $\dot{R}$ and obtain
\m
\label{dotR}
\dot{R}=\sqrt{\frac{2}{\mu}} \sqrt{E-\frac{\tilde{L}^{2}}{2 \mu R^{2}}-\frac{C}{R}}.
\n
From the definition, the conserved module of the real angular momentum can be expressed by
\m
\label{tildeL}
\tilde{L}\equiv \mu|\bm{R}\times \bm{v}|=\mu R v_{\perp}=\mu R^2 \sin \theta \dot{\phi},
\n
while the conserved module of the generalized angular momentum is given by
\m
\label{L}
L=\frac{\tilde{L}}{\sin \theta}=\mu R^2   \dot{\phi}.
\n
For the equation of the orbit, we need the relationship of $R$ and $\phi$ by eliminating the parameter $t$. From Eqs.~\eqref{dotR} and \eqref{L}, we get
\m
\frac{\dot{\phi}}{\dot{R}}=\frac{d \phi}{d R}=\left(\frac{2 \mu E}{L^{2}} R^{4}-\frac{2 \mu C}{L^{2}} R^{3}-\sin ^{2} \theta R^{2}\right)^{-\frac{1}{2}}.
\notag \\
\n

\begin{figure}[htpb]
    \includegraphics[width=0.48\textwidth]{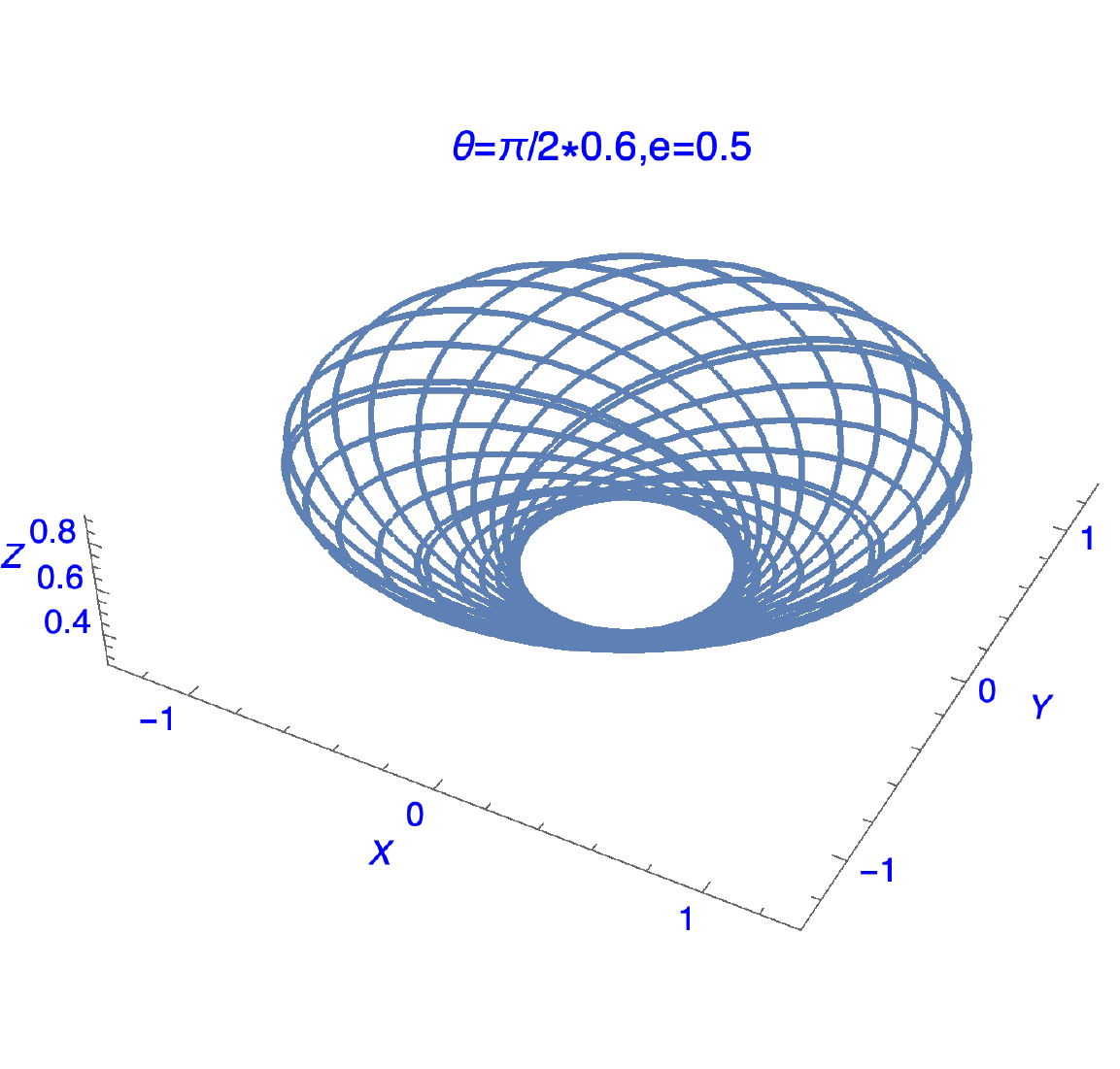}
    \caption[]{ A conic-shaped orbit of the binary which is confined to the surface of a cone is plotted in the range of $\phi$ from $0$ to $40 \pi/\sin \theta$ and the parameters $a=1,\, e=0.5$ and $\theta=\pi/2\times 0.6$ according to \eqref{R}.  Though the orbit is bounded, it is not closed and has an infinite period in three dimensions since $\sin(3 \pi/10) = (1+ \sqrt{5})/4$ is an irrational number.}
    \label{fig:R}
\end{figure}

Setting $x=1/R$ and integrating by quadrature,
we finally obtain the Keplerian orbit on the cone with half-aperture angle $\theta$
\m
R&=&\frac{\frac{\tilde{L}^{2}}{\mu|C|}}{1+\sqrt{1+\frac{2 \tilde{L}^{2}}{\mu C^{2}} E} \cos (\phi \sin \theta)}
\notag \\
&\equiv& \frac{a\left(1-e^{2}\right)}{1+e \cos (\phi \sin \theta)},
\n
while the conserved energy and the magnitude of angular momentum  are, respectively,
\m
E=\frac{C}{2a},
\n
and
\m
\label{tilde}
\tilde{L}^{2}=\mu|C| a\left(1-e^{2}\right).
\n
 Note that by introducing the conic section parameters $(a, e)$, we have expressed the radius on a conic section in the standard Keplerian form.  The bounded motion of our binary system requires $E<0$, which means $C<0$. From Eqs.~(\ref{tildeL}) and (\ref{tilde}), we obtain the evolution of the azimuthal angle
\m
\label{dphidt}
\dot{\phi}=\frac{(-C)^{\frac{1}{2}} \csc (\theta ) (e \cos (\phi  \sin (\theta ))+1)^2}{a^{\frac{3}{2}} \left(1-e^2\right)^{\frac{3}{2}} \mu^{\frac{1}{2}} }.
\n
Further, by choosing $z$-axis along $\bm{L}$, the orbit is explicitly given by
\m
\label{R}
\boldsymbol{R}=\frac{a\left(1-e^{2}\right)}{1+e \cos (\phi \sin \theta)}\left(\begin{array}{c}
\sin \theta \cos \phi \\
\sin \theta \sin \phi \\
\cos \theta
\end{array}\right).
\n
Note that $\theta$ is a constant, we are able to interpret Eq.~\eqref{R} as a conic-shaped orbit of the binary which is confined to the surface of a cone with half-aperture angle $\theta$, as shown in Fig.~\ref{fig:R}. In Fig.~\ref{fig:R}, we plot the orbit by choosing $a=1, \theta=\pi/2\times 0.6$ and $e=0.5$. Now that we have orbits of dyonic binaries, we have explored the features in general and chaotic behaviors of orbits in Appendix.

\begin{figure}[htpb]
    \includegraphics[width=0.4\textwidth]{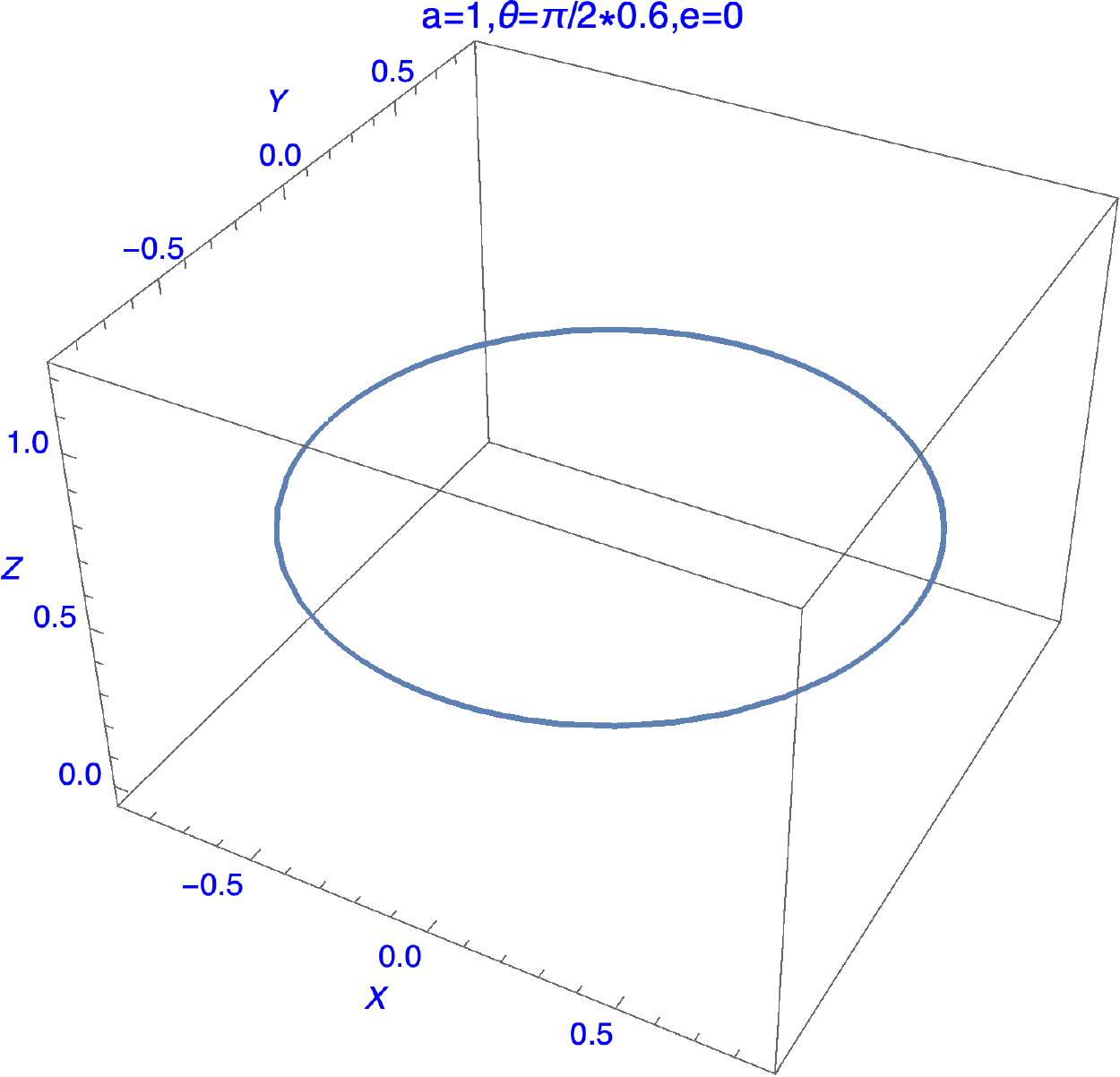}
    \caption[]{A circle-shaped orbit of the binary is plotted for the parameters $a=1,\, \theta=\pi/2\times 0.6$ and $ e=0$, according to Eq.~\eqref{Re=0}.  Though $\sin(3 \pi/10)$ is an irrational number, the orbit is closed, has a Keplerian form and has a finite period in three dimensions.}
   \label{fig:e=0}
\end{figure}

\begin{figure}[htpb]
     \includegraphics[width=0.4\textwidth]{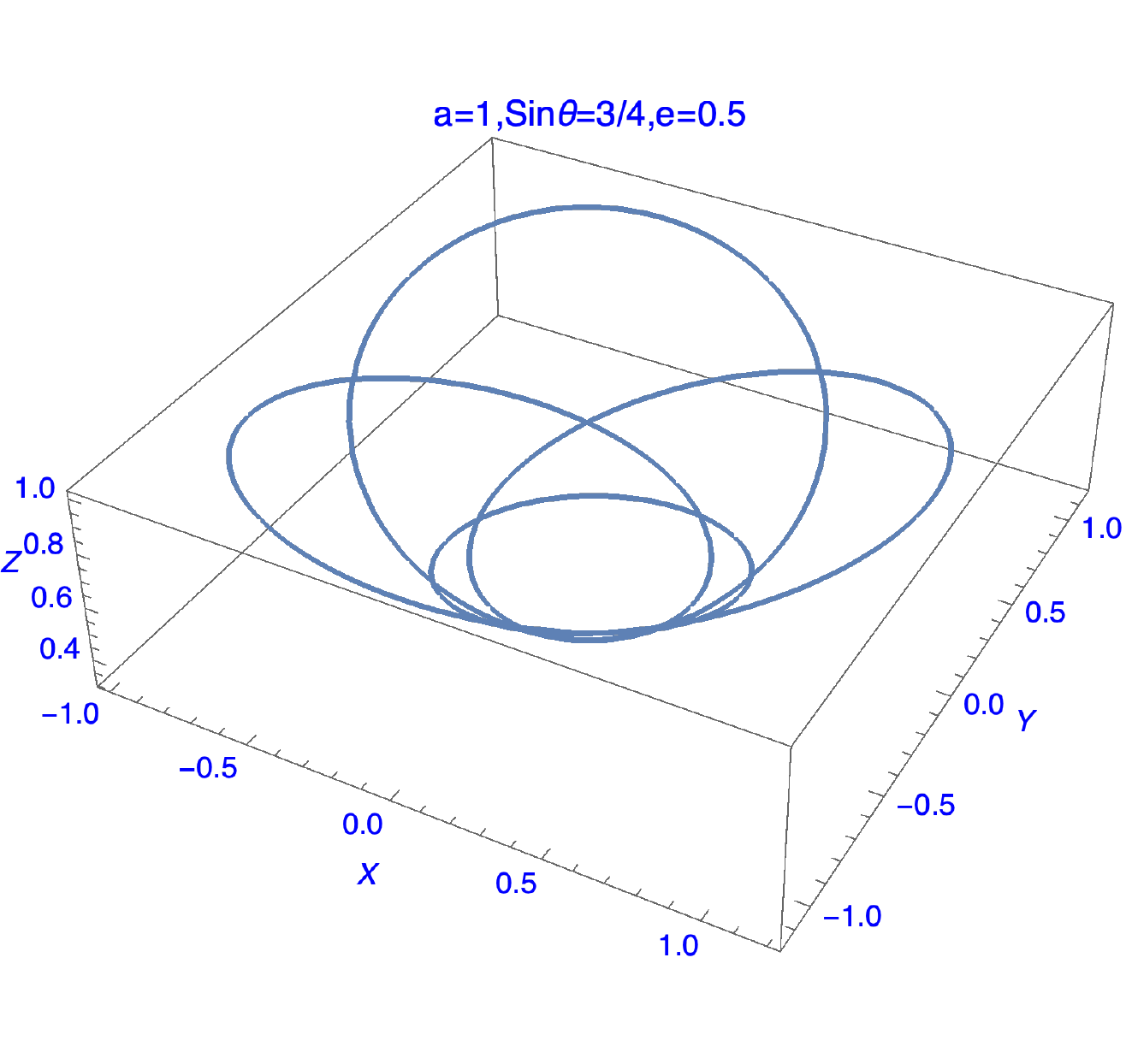}\\
     \includegraphics[width=0.4\textwidth]{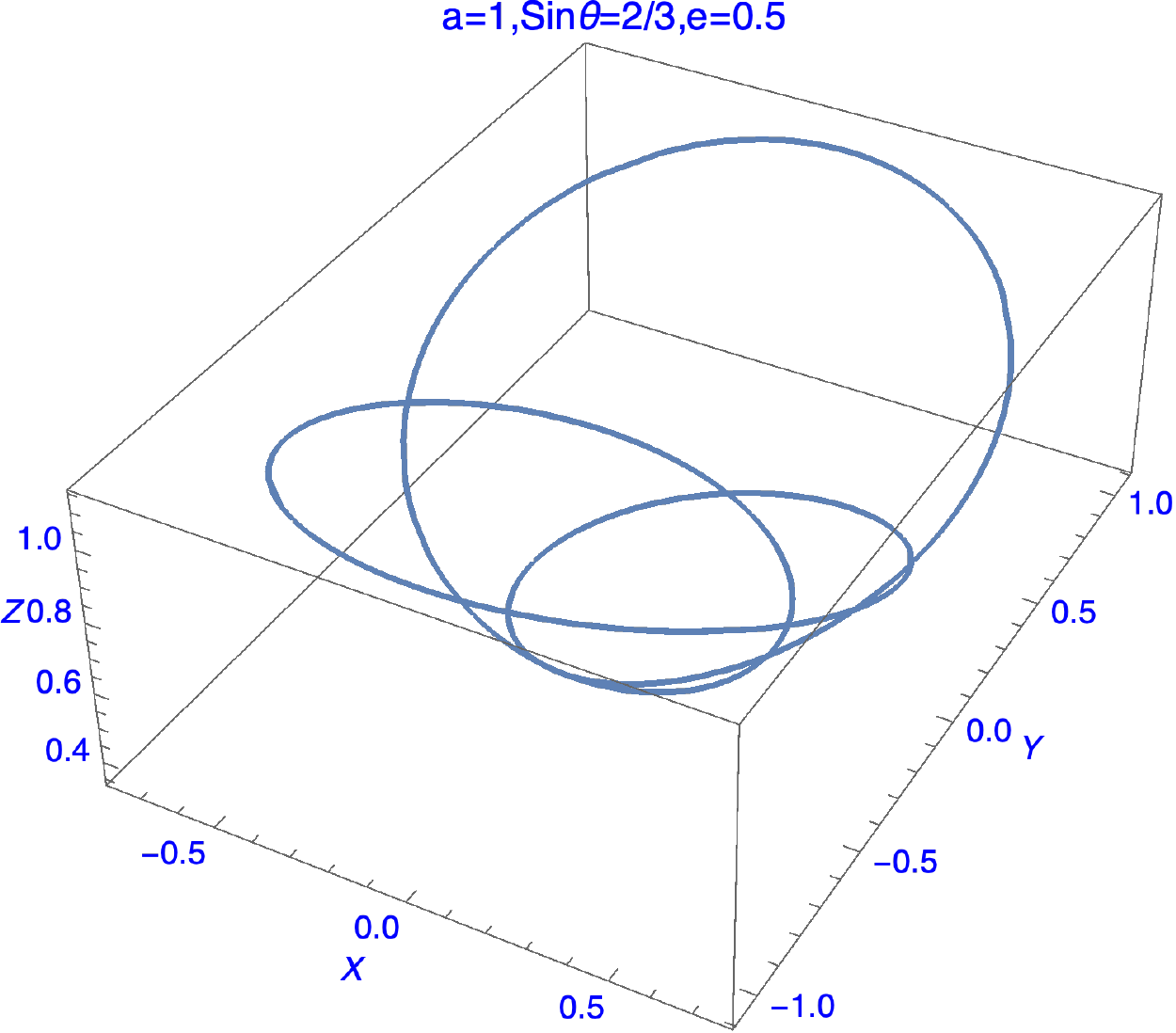}
    \caption[]{ Two different closed orbits of the binary with the parameters $a=1 $ and $e=0.5$ are illustrated for the rational values of  $\sin \theta=3/4$ (top) and $\sin \theta=2/3$ (bottom).}
    \label{fig:R4}
\end{figure}

First, let us consider the case $e=0$ for our binary system. From Eqs.~\eqref{dphidt} and \eqref{R}, the three-dimensional trajectory
\m
\label{Re=0}
\boldsymbol{R}=a\left(\begin{array}{c}
\sin \theta \cos \phi \\
\sin \theta \sin \phi \\
\cos \theta
\end{array}\right),
\n
is effectively a two-dimensional circular orbit with $z = \cos \theta$, as illustrated in Fig.~\ref{fig:e=0}, and the orbital rate
\m
\label{dphidte=0}
\dot{\phi}=\frac{(-C)^{\frac{1}{2}}}{\mu^{\frac{1}{2}} a^{\frac{3}{2}} \sin \theta },
\n
has a finite period
\m
 T_1=\int_0 ^{2\pi} d\phi \dot{\phi}^{-1}=2 \pi a^{3 / 2} \sqrt{-\mu / C} \sin \theta.
\n

Next, we consider a conical elliptical orbit of $e \neq 0$ for our binary system. To get a closed orbit, we need to analyze Eq.~\eqref{R}. If and only if $\sin \theta$ is a rational number
\m
\sin \theta=\frac{l}{n}
\n
with $l$ and $n$ relatively positive prime numbers and $l<n$, the orbit will be closed after $n$ revolutions, the system will complete one exact ellipse and return to the initial position. In such a case, one period is given by
\m
T_2=\int_0 ^{2n\pi} d\phi \dot{\phi}^{-1}=2 \pi a^{3 / 2} \sqrt{-\mu / C} l.
\n
Moreover, the different numbers $l$ and $n$ will determine the different topology of the orbit, as shown in Fig.~\ref{fig:R4}.

\begin{figure}[htpb]
    \includegraphics[width=0.48\textwidth]{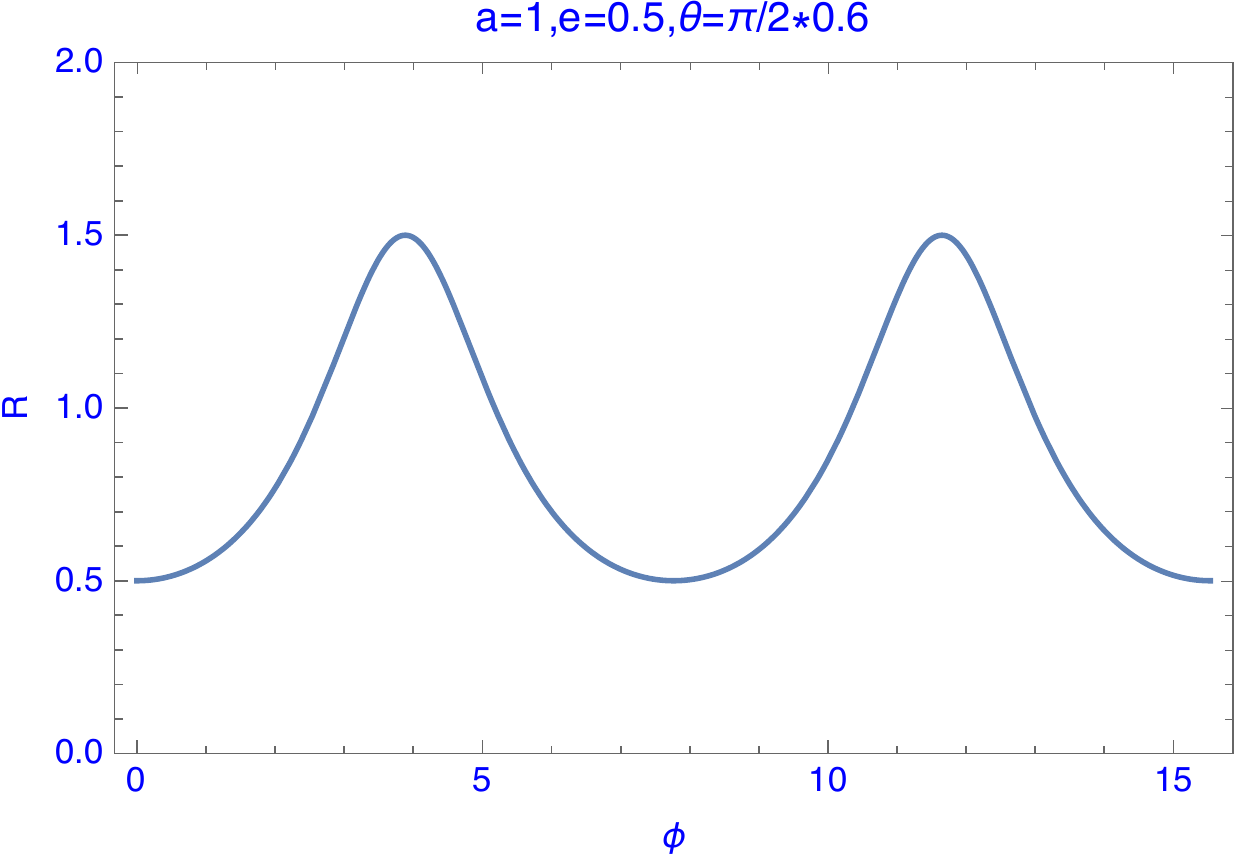}
    \caption[]{The plot of $R$ as a function of $\phi$ by choosing $a=1, \theta=\pi/2\times 0.6$ and $e=0.5$, according to \eqref{R}.}
    \label{fig:absR}
\end{figure}

For $e \neq 0$, no matter how rational or irrational $\sin \theta$ is, $R$ is a periodic function of $\phi(t)$ with the period
\m
 T_3=\int_0 ^{2\pi/\sin(\theta)} d\phi \dot{\phi}^{-1}=2 \pi a^{3 / 2} \sqrt{-\mu / C},
\n
as shown in Fig.~\ref{fig:absR}. When $e \neq 0$ and $\sin \theta$ is irrational, the orbit is not closed and shows a chaotic behavior of a conserved autonomous system \cite{Argyris:2015311}; for instance,  in Fig.~\ref{fig:R},  we plot the orbit of the binary by choosing $a=1, \theta=\pi/2\times 0.6$ and $e=0.5$.

Particularly when $|D|\ll L$, we have
\m
\cos \theta=|D| / L=\left(1+\frac{|C|}{D^{2}} \mu a\left(1-e^{2}\right)\right)^{-1 / 2},
\n
\m
\sin \theta \simeq 1,
\n
and the orbit is approximately given by
\m
\boldsymbol{R}\simeq \frac{a\left(1-e^{2}\right)}{1+e \cos (\phi )}\left(\begin{array}{c}
 \cos \phi \\
 \sin \phi \\
\left(1+\frac{|C|}{D^{2}} \mu a\left(1-e^{2}\right)\right)^{-1 / 2}
\end{array}\right).
\notag \\
\n
In the limiting case of $D=0$, the orbit becomes the Keplerian on the equatorial plane $(\theta = \pi/2)$.

Now that we have a description for orbits, we will calculate the emissions of energy and angular momentum for the $e=0$ case in the next section.

\section{Solutions with radiation}
\label{Emission}

In this section, we will consider the case of $e=0$ only, i.e, the circular orbits and leave the case of $e \neq 0$ for future work.  By using the quasi-static approximation\footnote{The quasi-static approximation is that  the emission is assumed constant during one averaging period.}, we will calculate the total emission rate of energy and angular momentum  due to  gravitational and electromagnetic radiations.

\subsection{Electromagnetic radiation}

We first calculate the emission of electromagnetic radiation from electric charges on the orbit \eqref{Re=0}, averaged over an orbital period. Then we consider the emission from magnetic charges in the same orbit and finally superimpose their fields. This derivation follows the same procedure as that in Refs.~\cite{Liu:2020cds,Christiansen:2020pnv}.

Following \cite{Liu:2020cds,Christiansen:2020pnv},
the vector potential $A$ at $\bm{r}$ $(r\gg a)$ is given by
\m
\label{A}
A^{i} \simeq \frac{P^{i j}}{\sqrt{4 \pi} r} \dot{Q}^{j},
\n
where
\m
Q^{i}=q_{1} x_{1}^{i}+q_{2} x_{2}^{i}=\mu \Delta \sigma_q R^{i},
\n
 is the electric dipole, $P^{i j}=\delta^{i j}-n^{i} n^{j}$ is the transverse projection, and
\m
 \Delta \sigma_q=q_{2} / m_{2}-q_{1} / m_{1}.
\n
 The energy emission due to electric charges is given by
 \m
 P_e=\frac{2 \mu^{2}(\Delta \sigma_q)^{2}}{3} \ddot{R}^i \ddot{R}_i,
 \n
and the average energy loss over an orbital period $T$ is
\m
\bar{P}_e&=&\frac{1}{T_1} \int_0 ^{2\pi} d\phi P_e \dot{\phi}^{-1}
\notag \\
&=&\frac{2 C^2 (\Delta \sigma_q)^{2} \csc ^2(\theta )}{3 a^4}.
\n
The angular momentum emission due to electric charges is given by
\m
\dot{J}_e^{i}=-\epsilon^{i}_{j k} \frac{2}{3}\dot{Q}^{j} \ddot{Q}^{k}=-\frac{2 \mu^{2}(\Delta \sigma_q)^{2}}{3} \epsilon_{j k}^{i}\dot{R}^{j} \ddot{R}^{k}.
\n
For the angular momentum loss due to electromagnetic radiation averaged one orbital period $T$, we have
\m
\left\langle\frac{dJ^i_e}{dt}\right\rangle \equiv \frac{1}{T_1} \int_{0}^{T_1} d t \dot{J^i_e}.
\n
For $e=0$, we can get
\m
\dot{J}_e^1=\dot{J}_e^2=\left\langle\dot{J}_e^1\right\rangle=\left\langle\dot{J}_e^2\right\rangle=0,
\n
\m
\left\langle\dot{J}_e^3\right\rangle=\dot{J}_e^3=-\frac{2  (-C)^{3/2} \sqrt{\mu} (\Delta \sigma_q)^{2} \csc (\theta )}{3 a^{5/2}}.
\n

A great consequence of the enhanced symmetry due to the existence of magnetic monopoles is that Maxwell's equations and thus the classical dynamics of all the fields and charges remain invariant under the dual transformation
\m
\begin{aligned}
&\boldsymbol{E}^{\prime}=\boldsymbol{E} \cos \alpha-\boldsymbol{B} \sin \alpha, \\
&\boldsymbol{B}^{\prime}=\boldsymbol{E} \sin \alpha+\boldsymbol{B} \cos \alpha,\\
&q^{\prime} = q \cos \alpha+g \sin \alpha,\\
&g^{\prime} = g \cos \alpha-g \sin \alpha.
\end{aligned}
\n
 By choosing $\alpha=\pi / 2$, pure electric charges transform to pure magnetic charges. This allows us to immediately find the fields emanating from magnetic charges on the orbit from our results for pure electric charges so far. We may then superimpose them to find the total emission.  For $\alpha=\pi / 2$, we see
\begin{equation}
\begin{aligned}
&\boldsymbol{E}_{2}=-\left(\frac{\Delta \sigma_{g}}{\Delta \sigma_{e}}\right) \boldsymbol{B}_{1},\\
&\boldsymbol{B}_{2}=+\left(\frac{\Delta \sigma_{g}}{\Delta \sigma_{e}}\right) \boldsymbol{E}_{1},
\end{aligned}
\end{equation}
where $\Delta \sigma_g$ determines the magnetic dipole:
\m
\Delta \sigma_g=g_{2} / m_{2}-g_{1} / m_{1}.
\n
Here, we have used the transformation between the electric and magnetic fields and the vector potential \eqref{A} to infer their proportionality with the charges.

In fact, we need not superimpose the fields. Instead we may superimpose their emissions.  This does not hold true in general since the superposition principle of the Maxwell theory applies only to the fields. However,  we show in the following that the addition of emissions applies for this specific situation of adding dual fields together.  A way to immediately and conceptually show that this is correct, is to notice that, far away on a shell where we calculate the emissions, the electric and magnetic fields from the electric charges are perpendicular to those from the magnetic charges, while the electric (magnetic) fields from the electric charges are parallel with the magnetic (electric) fields from the magnetic charges, and so, all the cross-terms vanish while superimposing the fields.

We explicitly show this by considering the integrated energy- and angular momentum-density on a shell for electric and magnetic fields, labeling the fields from the electric charge configuration $E_{1}, B_{1},$ and those from the dual transformation $E_{2}, B_{2}$.
Notice that the electric dipole has the same direction as the magnetic dipole. So, we have $\boldsymbol{E}_{1} \perp \boldsymbol{E}_{2}, \boldsymbol{B}_{1} \perp \boldsymbol{B}_{2}$, $\boldsymbol{E}_{1} ||\boldsymbol{B}_{2}$, $\boldsymbol{E}_{2} ||\boldsymbol{B}_{1}$. Now we look at the result for the energy density and momentum density:
\m
u&=&\frac{1}{2}\left(E^{2}+B^{2}\right)
=\frac{1}{2}(E_{1}^{2}+B_{1}^{2}
\notag \\
&+& E_{2}^{2}+B_{2}^{2}+2\left(\boldsymbol{E}_{1} \cdot \boldsymbol{E}_{2}+\boldsymbol{B}_{1} \cdot \boldsymbol{B}_{2}\right))=u_1+u_2.
\n
\m
\boldsymbol{\mathcal{P}}&=&\boldsymbol{E} \times \boldsymbol{B}=\boldsymbol{E}_{1} \times \boldsymbol{B}_{1}+\boldsymbol{E}_{2} \times \boldsymbol{B}_{2}
\notag \\
&+&\boldsymbol{E}_{1} \times \boldsymbol{B}_{2}+\boldsymbol{E}_{2} \times \boldsymbol{B}_{1}=\boldsymbol{\mathcal{P}_1}+\boldsymbol{\mathcal{P}_2}.
\n
Noting that
\m
P=- r^2 \int d\Omega \hat{r}\cdot \boldsymbol{\mathcal{P}},
\n
\m
\dot{J}=-r^{2}  \int d \Omega \boldsymbol{r} \times \boldsymbol{\mathcal{P}},
\n
we have
\m
\bar{P}_{EM}=\left(1+\left(\frac{\Delta \sigma_{g}}{\Delta \sigma_{e}}\right)^{2}\right) \bar{P}_1,
\n
\m
\dot{J}_{EM}=\left(1+\left(\frac{\Delta \sigma_{g}}{\Delta \sigma_{e}}\right)^{2}\right) \dot{J}_1.
\n
This means that our final results for the energy and angular momentum emissions from our binary system are
\m
\bar{P}_{EM}=\frac{2 C^2 ((\Delta \sigma_q)^{2}+(\Delta \sigma_g)^{2}) \csc ^2(\theta )}{3 a^4},
\n

\m
\left\langle\dot{J}_{EM}\right\rangle=-\frac{2  (-C)^{3/2} \sqrt{\mu} ((\Delta \sigma_q)^{2}+(\Delta \sigma_g)^{2}) \csc (\theta )}{3 a^{5/2}}.
\nonumber\\
\n
Following \cite{Liu:2020cds}, the gravitational field or electromagnetic field carries away a total angular momentum $J$, which consists of an orbital angular momentum contribution and  a spin contribution. This total angular momentum is drained from the total angular momentum of the source, which, for our binary system  has the origin of a pure orbital motion. So, the loss rates of the energy and angular momentum in our binary system due to electromagnetic radiation are given by
\m
\left\langle\frac{dE_{EM}}{dt}\right\rangle=-\frac{2 C^2 ((\Delta \sigma_q)^{2}+(\Delta \sigma_g)^{2}) \csc ^2(\theta )}{3 a^4},
\n
\m
\left\langle\frac{dL_{EM}}{dt}\right\rangle=-\frac{2  (-C)^{3/2} \sqrt{\mu} ((\Delta \sigma_q)^{2}+(\Delta \sigma_g)^{2}) \csc (\theta )}{3 a^{5/2}},
\notag \\
\n
from which we find the relation
\m
\frac{\left\langle\frac{dL_{EM}}{dt}\right\rangle }{\left\langle\frac{dE_{EM}}{dt}\right\rangle }= \sqrt{(-C)^{-1} \mu a^3} \sin(\theta) = \mu a^2/L.
\n

\subsection{Gravitational radiation}
Now, we compute the total radiated power in GWs.  In our reference frame where $\bm{L}$ is along $z$ axis, the second mass moment can be written as
\m
\label{Mij}
M^{ij}=\mu R^i R^j.
\n

Following \cite{Peters:1963ux}, the radiated power of GWs is expressed as
\m
P_{GW}=\frac{1}{5 }\left\langle\ddot{M}_{i j} \ddot{M}_{i j}-\frac{1}{3}\left(\ddot{M}_{k k}\right)^{2}\right\rangle
\n
Using Eqs.~\eqref{dphidte=0} and \eqref{Mij}, one has
\m
\begin{aligned}
&\dddot{M}_{11}=\frac{4 (-C)^{3/2} \csc (\theta ) \sin (2 \phi )}{a^{5/2} \sqrt{\mu }},
\\
&\dddot{M}_{12}=-\frac{4 (-C)^{3/2} \csc (\theta ) \cos (2 \phi )}{a^{5/2} \sqrt{\mu }},
\\
&\dddot{M}_{13}=\frac{(-C)^{3/2} \cot (\theta ) \csc (\theta ) \sin (\phi )}{a^{5/2} \sqrt{\mu }},
\\
&\dddot{M}_{22}=-\frac{8 (-C)^{3/2} \csc (\theta ) \sin (\phi ) \cos (\phi )}{a^{5/2} \sqrt{\mu }},
\\
&\dddot{M}_{23}=-\frac{(-C)^{3/2} \cot (\theta ) \csc (\theta ) \cos (\phi )}{a^{5/2} \sqrt{\mu }},
\\
&\dddot{M}_{33}=0.
\end{aligned}
\n
Therefore, we obtain
\m
P_{GW}=-\frac{(-C)^3 (15 \cos (2 \theta )-17) \csc ^4(\theta )}{5 a^5 \mu },
\n
which is independent of $\phi$. The energy of GWs is only well-defined by taking an average over one period. In our case, a well-defined quantity is the average of $P_{GW}$ over one period $T_1$. Thus we perform the time average to get the total radiated power
\m
\bar{P}_{GW} \equiv \frac{1}{T_1} \int_{0}^{T_1} d t P_{GW}=P_{GW},
\n
and the averaged energy loss over an orbital period $T_1$ is given by
\m
\left\langle\frac{dE_{GW}}{dt}\right\rangle=-\bar{P}_{GW}=\frac{(-C)^3 (15 \cos (2 \theta )-17) \csc ^4(\theta )}{5 a^5 \mu }.
\notag \\
\n

Similarly, following \cite{Peters:1964zz}, the rate of angular momentum emission due to GWs is given by
\m
\frac{d L^{i}_{GW}}{d t}=-\frac{2}{5} \epsilon^{i k l}\left\langle\ddot{M}_{k a} \ddot{M}_{l a}\right\rangle.
\n
We have the angular momentum loss due to gravitational radiation averaged one orbital period $T_1$:
\m
\left\langle\frac{dL^i_{GW}}{dt}\right\rangle \equiv \frac{1}{T_1} \int_{0}^{T_1} d t \dot{L}^i_{GW}.
\n
For the case of $e=0$, i.e.,  circular orbits, we get
\m
\dot{L}_{GW}^1=\frac{12 (-C)^{5/2} \cot (\theta ) \csc (\theta ) \cos (\phi )}{5 a^{7/2} \sqrt{\mu }},
\n
\m
\dot{L}_{GW}^2=\frac{12 (-C)^{5/2} \cot (\theta ) \csc (\theta ) \sin (\phi )}{5 a^{7/2} \sqrt{\mu }},
\n
from which  it follows
\m
\left\langle\dot{L}_{GW}^1\right\rangle=\left\langle\dot{L}_{GW}^2\right\rangle=0,
\n
and  the nonvanished average
\m
\left\langle\dot{L}_{GW}^3\right\rangle=\dot{L}_{GW}^3=\frac{(-C)^{5/2} (15 \cos (2 \theta )-17) \csc ^3(\theta )}{5 a^{7/2} \sqrt{\mu }}.
\notag \\
\n
Because $\bm{L}$ is along $z$ axis, we thus conclude that the loss rates of the energy and angular momentum in our binary system due to gravitational radiation are
\m
\left\langle\frac{dE_{GW}}{dt}\right\rangle=\frac{(-C)^3 (15 \cos (2 \theta )-17) \csc ^4(\theta )}{5 a^5 \mu },
\n
\m
\left\langle\frac{dL_{GW}}{dt}\right\rangle=\frac{(-C)^{5/2} (15 \cos (2 \theta )-17) \csc ^3(\theta )}{5 a^{7/2} \sqrt{\mu }},
\n
 the ratio
\m
\frac{\left\langle\frac{dL_{GW}}{dt}\right\rangle }{\left\langle\frac{dE_{GW}}{dt}\right\rangle }= \sqrt{(-C)^{-1} \mu a^3} \sin(\theta) = \mu a^2/L.
\n
From the results for electromagnetic and gravitational radiations, we can show, for any $\theta$ and $a$, a universal relation
\m
\label{E-L-ratio}
\frac{\left\langle\frac{dE_{GW}}{dt}\right\rangle}{\left\langle\frac{dE_{EM}}{dt}\right\rangle}  = \frac{\left\langle\frac{dL_{GW}}{dt}\right\rangle}{\left\langle\frac{dL_{EM}}{dt}\right\rangle}.
\n

\subsection{Evolutions of $a$ and $\theta$}

For the static orbit, $\theta$ and $a$ are constants. However, when the emissions of energy and angular momentum due to gravitational and electromagnetic radiations are included, $\theta$ and $a$ become functions of time $t$. In this subsection, we will explore the evolutions of $a$ and $\theta$.

The total emission rates of energy and angular momentum due to gravitational and electromagnetic radiations are given, respectively, by
\m
\left\langle\frac{dE}{dt}\right\rangle=\left\langle\frac{dE_{EM}}{dt}\right\rangle+\left\langle\frac{dE_{GW}}{dt}\right\rangle,
\n
\m
\left\langle\frac{dL}{dt}\right\rangle=\left\langle\frac{dL_{EM}}{dt}\right\rangle+\left\langle\frac{dL_{GW}}{dt}\right\rangle.
\n
From the definitions
\m
E=\frac{C}{2a},
\n
\m
\label{L-evol}
L=\frac{\sqrt{-\mu C a}}{\sin(\theta)},
\n
we find the evolution of the semimajor axis and the conic angle
\m
\label{dadt}
\frac{da}{dt}&=& \frac{4 C ((\Delta \sigma_q)^{2}+(\Delta \sigma_g)^{2}) \csc ^2(\theta )}{3 a^2}
\notag \\
&+& \frac{2 C^2 (15 \cos (2 \theta )-17) \csc ^4(\theta )}{5 a^3  \mu},
\n
\m
\label{dthetadt}
\frac{d\theta}{dt}&=& \frac{2 C (\Delta \sigma_q)^{2}+(\Delta \sigma_g)^{2}) \cot (\theta )}{3 a^3}
\notag \\
&+& \frac{C^2 (15 \cos (2 \theta )-17) \cot (\theta ) \csc ^2(\theta )}{5 a^4 \mu }.
\n
 Note that  these two expressions have a similar form. Using the chain rule for differentiation, we can  find
\m
\frac{da}{d\theta}=2 a \csc (\theta ) \sec (\theta ),
\n
\m
a=c_0 \tan ^2(\theta),
\n
where $c_0$ is determined by the initial condition $a=a_0$ when $\theta=\theta_0$.
From Eq.~(\ref{L-evol}) and
\m
\cos(\theta)= |D|/L,
\n
we have
\m
c_0=\frac{D^2}{-\mu C}.
\n
Now we get the evolutions of orbits by using $dL/dt$. When we consider gravitational radiation and electromagnetic radiation, we find only $\dot{L}^3 \neq 0$, which implies that the direction of $\bm{L}$ does not change and the magnitude of $\bm{L}$ decreases while the direction and magnitude of $\bm{\tilde{L}}$ both change. We show how to use $d\tilde{L}/dt$ to get the same results of $da/dt$ and $d\theta/dt$ in Appendix. Next, we calculate merger times of dyonic binaries. For  arbitrary  $\theta$ and $a$, Eq.~(\ref{E-L-ratio}) is explicitly given by
\m
\frac{\left\langle\frac{dE_{GW}}{dt}\right\rangle}{\left\langle\frac{dE_{EM}}{dt}\right\rangle}  = \frac{\left\langle\frac{dL_{GW}}{dt}\right\rangle}{\left\langle\frac{dL_{EM}}{dt}\right\rangle} = \frac{3 C (15 \cos (2 \theta )-17) \csc ^2(\theta )}{10 a ((\Delta \sigma_q)^{2}+(\Delta \sigma_g)^{2}) \mu }.
\notag \\
\n
 From Eq.~(\ref{dadt}) in Appendix, we find that the rates of the semimajor axis increase quickly when semimajor axis decreases no matter what the first term or the second term dominates. So, the binary system spends most of the decay time in the part of orbit where $a \approx a_0$. For a given $a_0$, the total rate of energy and angular momentum emission is dominated by gravitational radiation or electromagnetic radiation which depends on $m_1,m_2,q_1,q_2,g_1$ and $g_2$.

First, we consider the case where electromagnetic radiation dominates over the gravitational radiation around $a \approx a_0$.
When $D=0$ or $\theta=\pi/2$, we have the rates
\m
\frac{da}{dt}=\frac{4 C ((\Delta \sigma_q)^{2}+(\Delta \sigma_g)^{2})  }{3 a^2},
\n
\m
\frac{d\theta}{dt}=0,
\n
which gives the coalescence time
\m
\label{EMtheta=pi/2}
\tau_{EM}(a_0,\theta=\frac{\pi}{2})= \int_{a_{0}}^{0} da (\frac{da}{dt})^{-1}=\frac{-a_0^3}{4 C ((\Delta \sigma_q)^{2}+(\Delta \sigma_g)^{2}) }.
\notag \\
\n

On the other hand, when $D\neq0$ or $\theta\neq\pi/2$, Eqs.~(\ref{dadt}) and (\ref{dthetadt}) become
\m
\label{dadt1}
\frac{da}{dt}=\frac{4 C ((\Delta \sigma_q)^{2}+(\Delta \sigma_g)^{2})  \csc ^2(\theta )}{3 a^2},
\n
\m
\label{dthetadt1}
\frac{d\theta}{dt}=\frac{2 C ((\Delta \sigma_q)^{2}+(\Delta \sigma_g)^{2}) \cot (\theta )}{3 a^3}.
\n
So, we similarly have
\m
\label{dadtheta1}
\frac{da}{d\theta}=2 a \csc (\theta ) \sec (\theta )
\n
\m
a=c_0\tan ^2(\theta ).
\n
Then we can have 
\m
(\frac{d\theta}{dt})^{-1}=\frac{3 c_0{}^3 \tan ^7(\theta )}{2 C ((\Delta \sigma_q)^{2}+(\Delta \sigma_g)^{2})}.
\n
For the orbit, we can integrate Eq.~\eqref{dadt1} by requiring $a(t)=0$ at $t=\tau_{EM}\left(a_{0}, \theta_{0}\right)$ or, equivalently, we can integrate Eq.~\eqref{dthetadt1} by requiring $\theta(t)=0$ at $t=\tau_{EM}\left(a_{0}, \theta_{0}\right)$, since we can see that at the coalescence $\theta$ goes to zero. Because the analytic expression for $a(\theta)$ is simpler than the form of the inverse function $\theta(a)$, it is much better to use Eq.~\eqref{dadtheta1} and to get
\m
\int_{0}^{\tau_{EM}\left(a_{0}, \theta_{0}\right)}dt=\int_{\theta_{0}}^{0} d \theta (\frac{d\theta}{dt})^{-1} ,
\n

\begin{figure}[htpb]
    \includegraphics[width=0.48\textwidth]{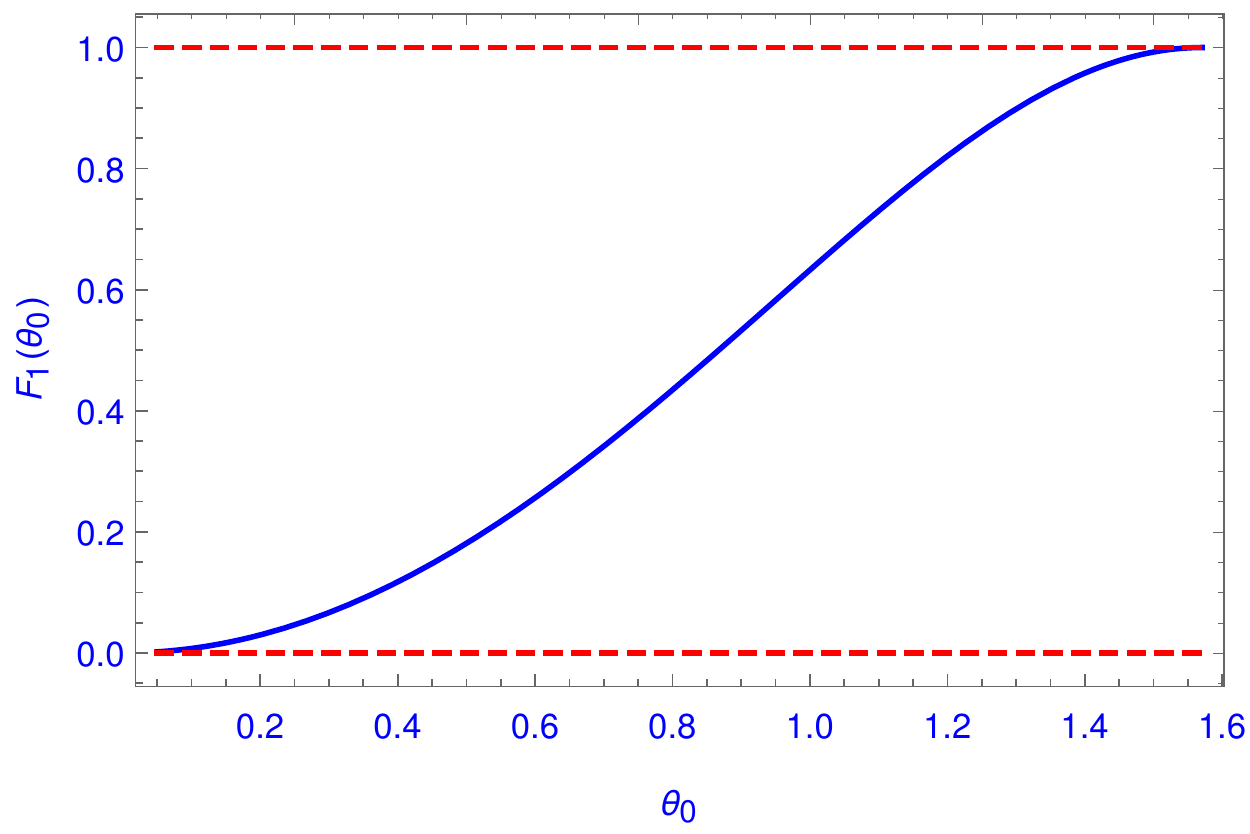}
    \includegraphics[width=0.48\textwidth]{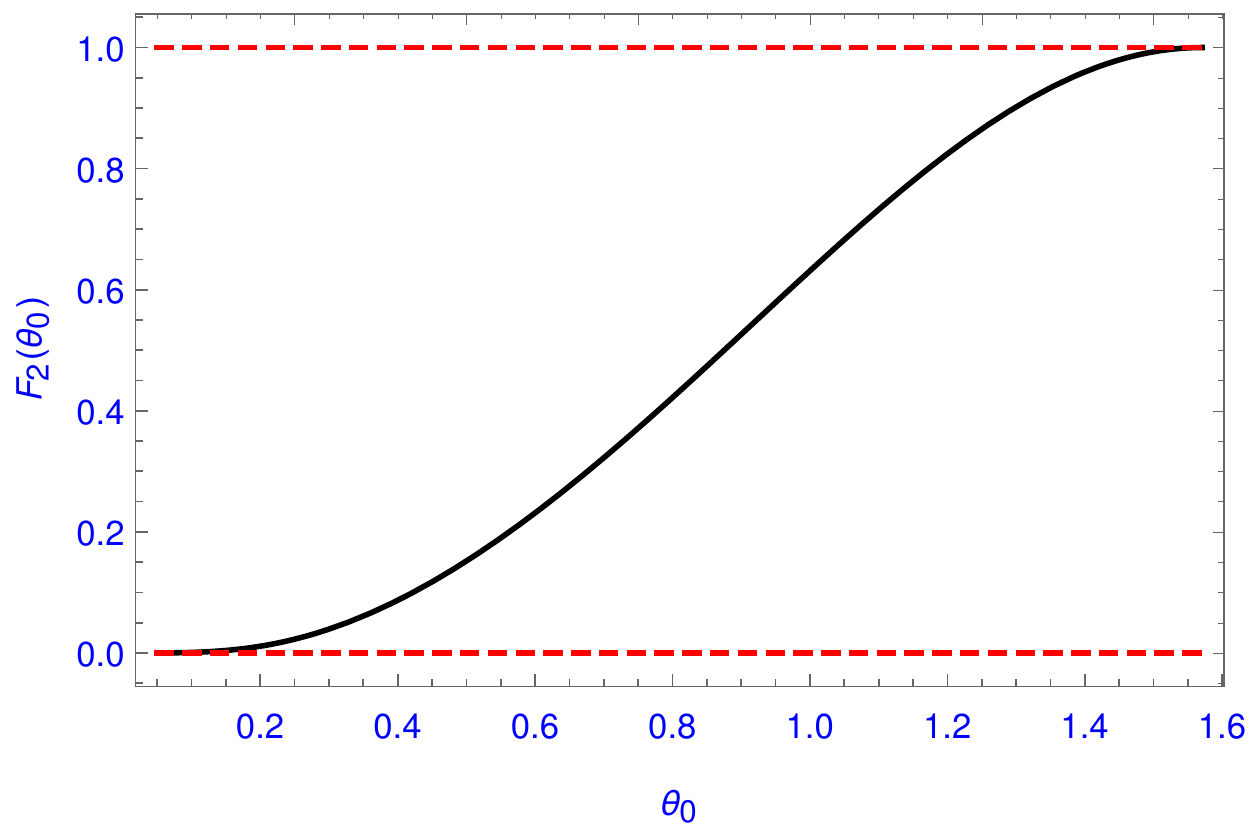}
    \caption[]{Upper panel: The plot of $F_1(\theta_0)$ as a function of $\theta_0$. Bottom panel: the plot of $F_2(\theta_0)$ as a function of $\theta_0$.}
    \label{fig:g1}
\end{figure}

\m
\tau_{EM}(a_0,\theta_0)&=&-\frac{ a_0^3 F_1(\theta_0)}{4 C ((\Delta \sigma_q)^{2}+(\Delta \sigma_g)^{2})},
\n
where
\m
\label{F1}
F_1(\theta_0)&=&
 \frac{1}{2\tan ^6(\theta_0) } (2 \sec ^6(\theta_0 )-9 \sec ^4(\theta_0 )+18 \sec ^2(\theta_0 )
\notag \\
&+&12 \log (\cos (\theta_0 ))-11),
\n
\m
\tan (\theta_0)=\frac{\sqrt{-\mu C a_0}}{|D|}.
\n
We plot $F_1(\theta_0)$ as a function of $\theta_0$ in Fig.~\ref{fig:g1}.  When $\theta_0\rightarrow 0$, $F_1(\theta_0) \rightarrow 0$ while $\theta_0\rightarrow \pi/2$, $F_1(\theta_0) \rightarrow 1$, which is consistent with Eq.~\eqref{EMtheta=pi/2} and that obtained  in  \cite{Liu:2020cds}.

Next, we consider the case where gravitational radiation dominates over the electromagnetic radiation around $a \approx a_0$.
When $D=0$ or $\theta=\pi/2$, we have 
\m
\frac{da}{dt}= \frac{ -64 C^2  }{5 a^3 \mu },
\n
\m
\frac{d\theta}{dt}=0,
\n
which  gives the merger time
\m
\label{GWtheta=pi/2}
\tau_{GW}(a_0,\theta=\frac{\pi}{2})= \int_{a_{0}}^{0} da (\frac{da}{dt})^{-1}= \frac{5a^4 \mu}{256 C^2}.
\notag \\
\n

In the other case where $D\neq0$ or $\theta\neq\pi/2$, Eqs.~\eqref{dadt} and \eqref{dthetadt} become
\m
\frac{da}{dt}= \frac{2 C^2 (15 \cos (2 \theta )-17) \csc ^4(\theta )}{5 a^3 \mu },
\n
\m
\frac{d\theta}{dt}=\frac{C^2 (15 \cos (2 \theta )-17) \cot (\theta ) \csc ^2(\theta )}{5 a^4 \mu }.
\n
 The functional dependence of $da/d \theta$ has the same form as Eq.~\eqref{dadtheta1}. Similarly, the coalescence time for the gravitational-radiation dominated merger is given by
\m
\tau_{GW}(a_0,\theta_0)&=&\frac{5a_0^4 \mu F_2(\theta_0)}{256 C^2},
\n
where
\m
\label{F2}
F_2(\theta_0)&=&\frac{1}{245760 \tan ^8(\theta_0 )}(245760 \sec ^8(\theta _0)
\notag \\
&-&1331200 \sec ^6\left(\theta _0\right)+3043200 \sec ^4(\theta _0)
\notag \\
&-&4124400 \sec ^2\left(\theta _0\right)-\log \left(15 \sin ^2\left(\theta _0\right)+1\right)
\notag \\
&-&2097150 \log \left(\cos \left(\theta _0\right)\right)+2166640).
\n

We plot $F_2(\theta_0)$ as a function of $\theta_0$ in Fig.~\ref{fig:g1}. When $\theta_0\rightarrow 0$, $F_2(\theta_0) \rightarrow 0$ while $\theta_0\rightarrow \pi/2$, $F_2(\theta_0) \rightarrow 1$, which is again consistent with Eq.~\eqref{EMtheta=pi/2} and that found in  \cite{Liu:2020cds}. From Eqs.~\eqref{F1} and \eqref{F2}, when $\theta_0$ goes near zero, $F_1(\theta_0)$ and $F_2(\theta_0)$ become much smaller than unit (\textit{e.g.~}$F_1(\pi/20)=0.018$, $F_2(\pi/20)=0.005$), which implies the electric and magnetic charges or $\theta_0$ can significantly suppress merger times of the dyonic binaries.
 As $\theta_0$ approaches to zero, the binary coalesces immediately and $\tau_{EM}$ and $\tau_{GW}$ vanish, as expected. But this case corresponds to the general relativistic regime, which requires methods beyond the Newtonian method. In this paper, we have only considered the leading orders of orbits and radiations, namely, the 0-PN corrections. Some aspects of the higher PN corrections in the context of dyonic black holes will be discussed in the future.

\section{conclusions and discussions}\label{Concl}

 Dyonic black holes have attracted much attention not only in theoretical study but also in recent observations of GWs. In this paper, we have derived the equations of motion of dyonic binaries and explored features of the static orbit including the chaotic behavior. By using a Newtonian method with radiation reactions included, we have calculated the total emission rate of energy and angular momentum due to gravitational and electromagnetic radiations for circular orbits. Moreover, we have found the evolution of $a$, $\theta$ and calculated  merger times of dyonic binaries. It has been shown that the electric and magnetic charges can significantly suppress merger times of dyonic binaries no matter what gravitational radiation or electromagnetic radiation dominates. The results of this paper provide rich informations about dyonic binaries and may be used to test black holes with magnetic charges.

Finally, we would like to discuss the potential implications of our formulae. So far, many merger events of binary black holes have been reported by \lvc. Our methods can be applied to the early inspiral of low-mass binaries in LIGO (like GW151226), which have a long signal in the detector-band. However, when LIGO/Virgo black holes nearly merge, our methods are not valid and a higher-order PN expansion or numerical-relativity simulations are needed. In the future, many merger events are expected to be detected by space-based GW detectors, such as LISA and Taiji. Space-based GW detectors might be able to detect massive black hole binaries in the inspiral phase. The inspiral of such massive black hole binaries can last several days, months or even years in the frequency band of space-based GW detectors. Obviously, our formulae are applicable to those inspirals of massive black hole binaries. Such merger events will provide a good chance for our formulae to investigate whether these black holes indeed have electric and magnetic charges by using inspiral waveforms, which provides an unexplored arena to probe fundamental physics in the standard model of particle physics and beyond. We leave these topics for future studies.

\acknowledgments
We thank the anonymous referee for providing constructive comments and suggestions to improve the quality of this paper.
This work is supported in part by the National Natural Science Foundation of China Grants
 No.11690021, No.11690022, No.11851302, No.11821505, and No.11947302,
 in  part by the Strategic Priority Research Program of the Chinese Academy of Sciences Grant No. XDB23030100,
No. XDA15020701 and by Key Research Program of Frontier Sciences, CAS.
The work of S.P.K. was supported in part by National Research Foundation of Korea (NRF) funded by the Ministry of Education (2019R1I1A3A01063183).
\appendix
\section*{Appendix A: Chaotic Orbits}\label{Appen A}

\begin{figure}[htpb]
    \includegraphics[width=0.48\textwidth]{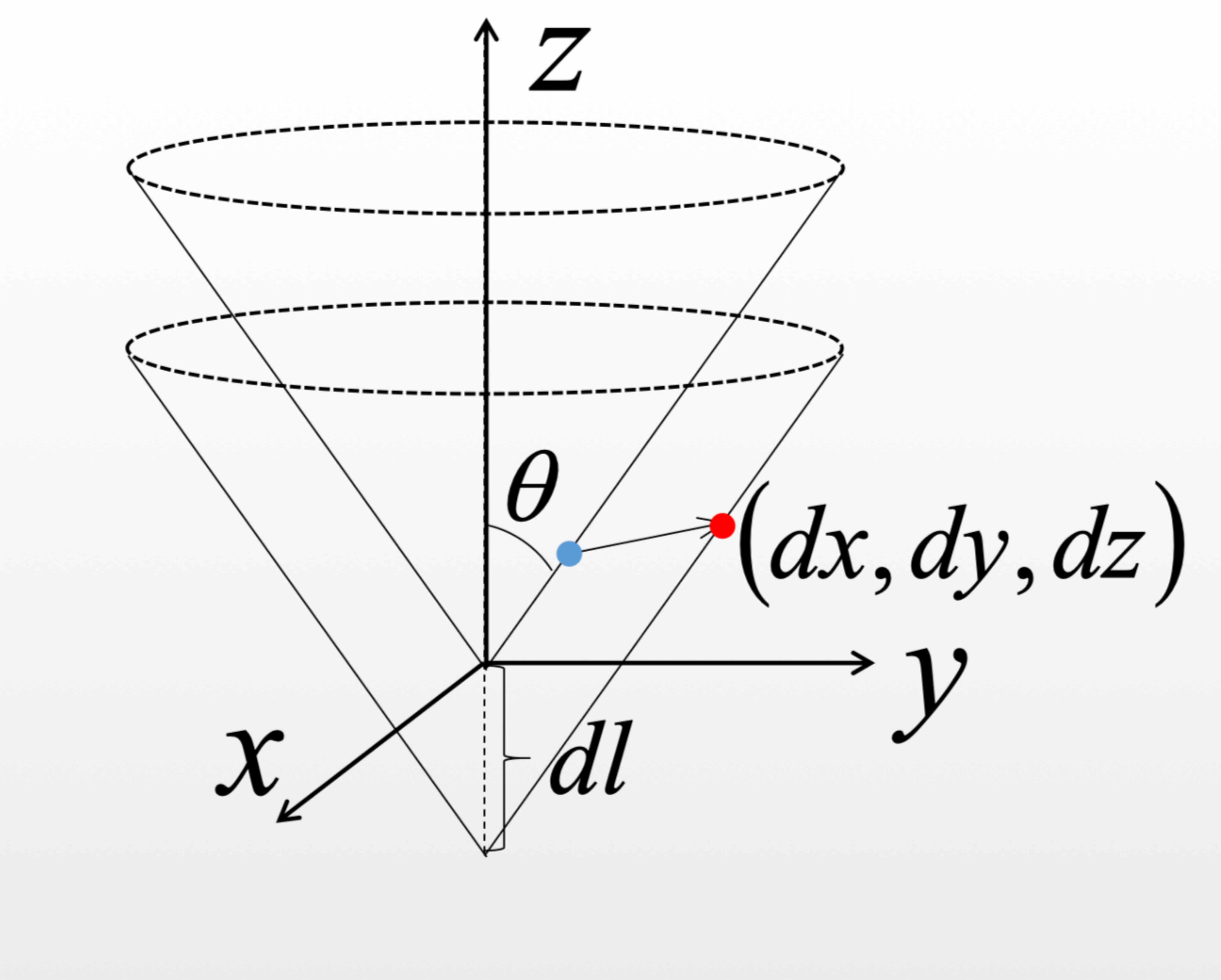}
    \caption[]{Schematic illustration on two neighboring orbits $\bm{R}(\phi)$ and $\bm{R_1}(\phi)$.}
    \label{fig:dl}
\end{figure}

Here we will explore chaotic behaviors of orbits by observing how two neighboring orbits deviate from each other during their evolutions. For this, we choose two neighboring orbits $\bm{R}(\phi)$ and $\bm{R_1}(\phi)$.
At time $t=0$ (assuming $\phi_i=0$), we set the initial separation as
\m
\bm{R_1}(0)-\bm{R}(0)=(dx, dy, dz).
\n
Notice that we just change the initial positions but the generalized angular momentum and the charges  $C$ and $D$ are fixed. 
And then we introduce a new parameter $dl$ which corresponds to the distance of two cones as shown in Fig.~\ref{fig:dl}. The orbit of $\bm{R_1}(\phi)$ can be expressed as
\m
\label{R1}
\boldsymbol{R_1}&=&\frac{a_1\left(1-e_1^{2}\right)}{1+e_1 \cos ((\phi+d\phi) \sin \theta)}\left(\begin{array}{c}
\sin \theta \cos (\phi+d\phi) \\
\sin \theta \sin (\phi+d\phi) \\
\cos \theta
\end{array}\right)
\notag \\
&-&dl \left(\begin{array}{c}
0 \\
0 \\
1
\end{array}\right),
\n
where
\m
a_1=a+da,  \quad  e_1=e+de.
\n
From the conserved magnitude of the real angular momentum,
\m
\tilde{L}^{2}=\mu|C| a\left(1-e^{2}\right)=\mu|C| a_1\left(1-e_1^{2}\right),
\n
we obtain the change of eccentricity,
\m
de=\frac{(1-e^2)da}{2ae}.
\n
To proceed further, we solve the vector equation
\m
\label{vetor}
&&\left.\frac{\partial \bm{R}}{\partial \phi}\right|_{\phi=0}d\phi+\left.\frac{\partial \bm{R}}{\partial a}\right|_{\phi=0}da+\left.\frac{\partial \bm{R}}{\partial e}\right|_{\phi=0}de
\notag \\
&-&dl \left(\begin{array}{c}
0 \\
0 \\
1
\end{array}\right)=\left(\begin{array}{c}
dx \\
dy \\
dz
\end{array}\right).
\n
Solving Eq.~\eqref{vetor}, we have
\m
da=-\frac{2 e \csc (\theta )}{\left(1-e\right)^2} dx,
\n
\m
d\phi=\frac{ \csc (\theta )}{a \left(1-e\right)}dy,
\n
\m
dl=\text{dx} \cot (\theta )-dz.
\n

Now, we introduce a new function $\lambda(\phi)$ defined as
\m
\label{la}
\lambda(\phi)\equiv\frac{(\bm{R_1}(\phi)-\bm{R}(\phi))^2}{(\bm{R_1}(0)-\bm{R}(0))^2},
\n
which describes the ratio of the evolution of the distance between two neighboring orbits $\bm{R}(\phi)$ and $\bm{R_1}(\phi)$.

\begin{figure}[htpb]
    \includegraphics[width=0.48\textwidth]{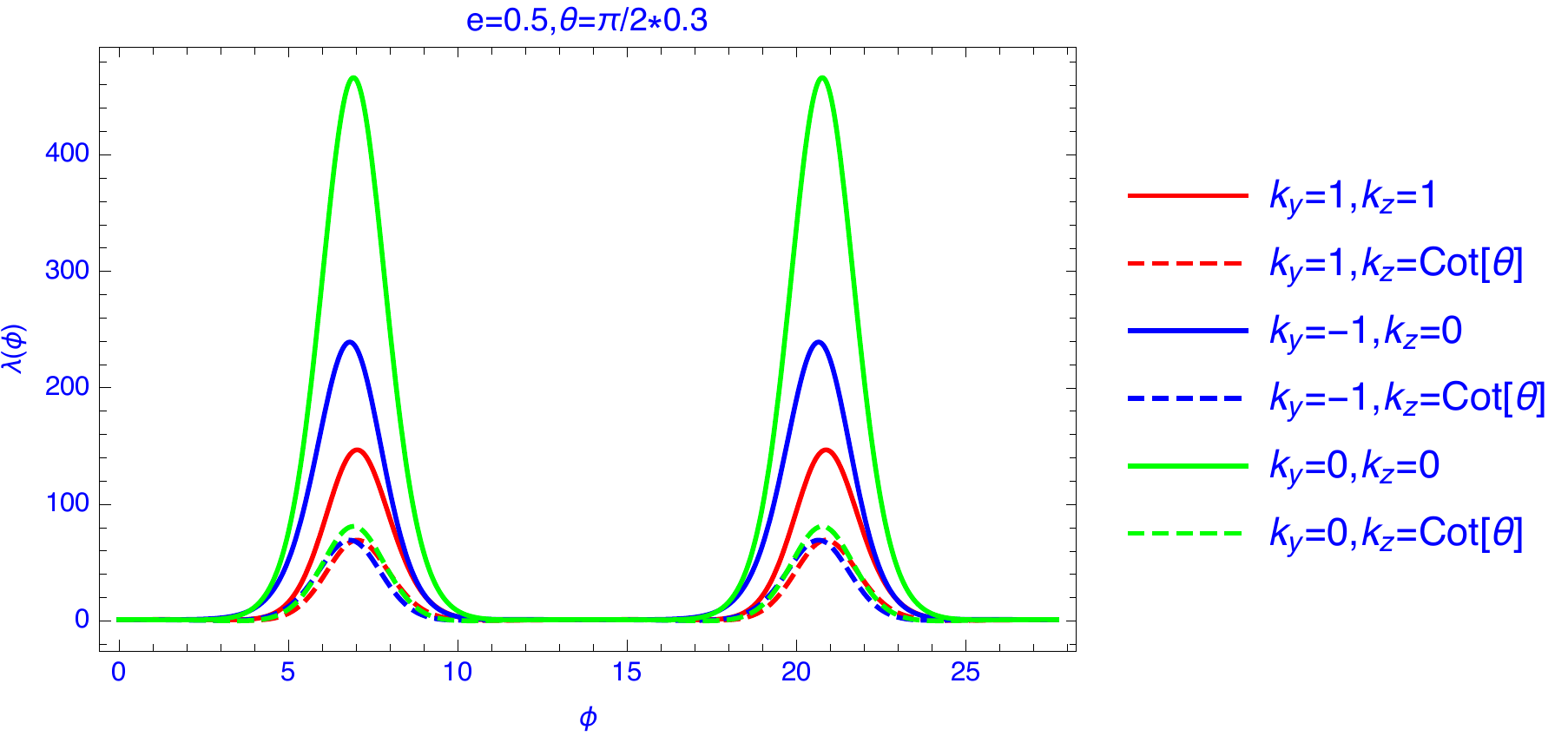}
    \includegraphics[width=0.48\textwidth]{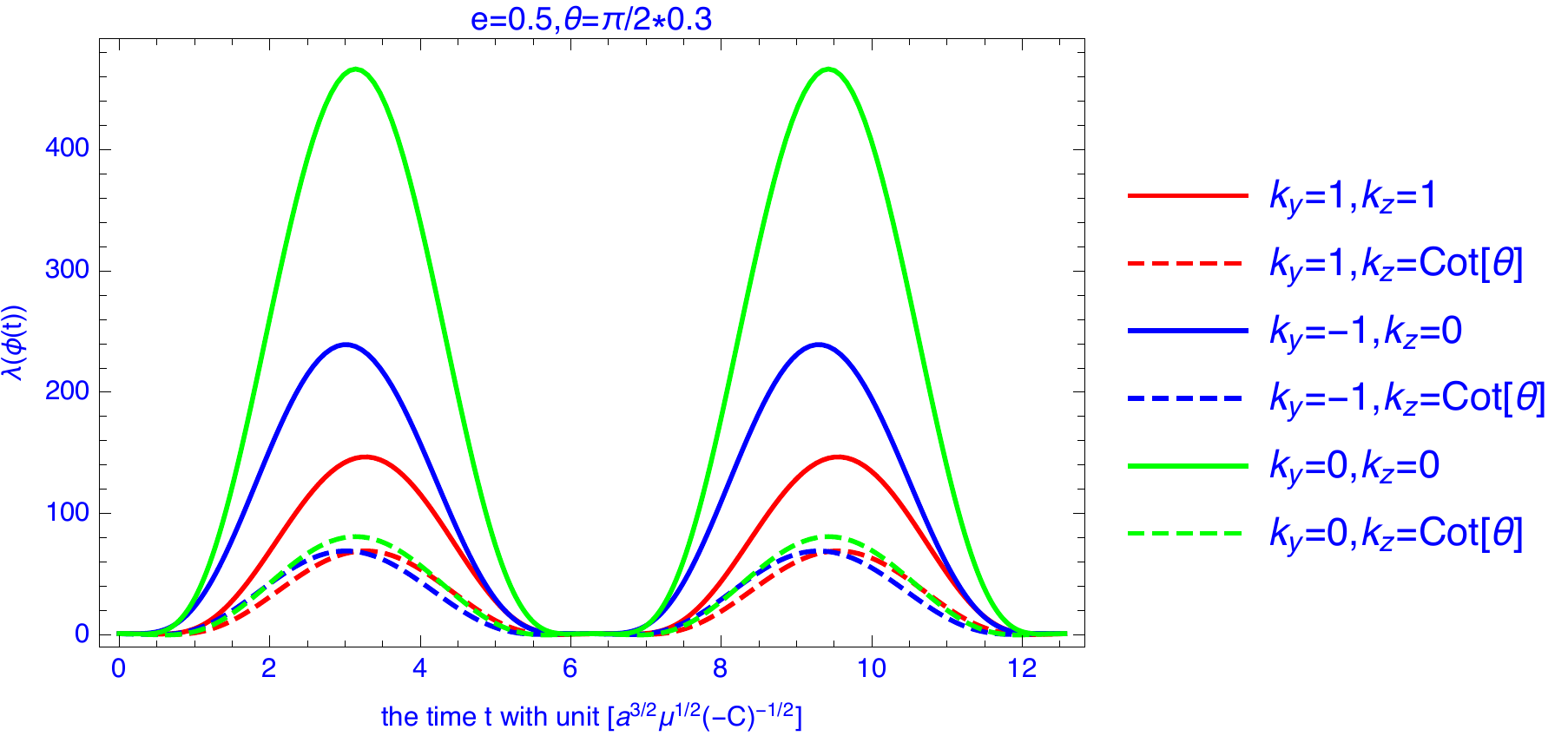}
    \caption[]{Upper panel: The plot of $\lambda (\phi)$ as a function of $\phi$ by choosing $ \theta=\pi/2\times 0.3$ and $e=0.5$, according to \eqref{laphi}. Bottom panel: The plot of $\lambda (\phi(t))$ as a function of $t$ by choosing $ \theta=\pi/2\times 0.3$ and $e=0.5$. Dashed lines represent the case $dl=0$ while solid lines represent the case $dl \neq 0$. For each period, there are growths (decreases) in time.}
    \label{fig:la}
\end{figure}

For the case, $dx \neq 0$, we can set
\m
\bm{R_1}(0)-\bm{R}(0)=(dx, dy, dz)=(1,k_y, k_z)\frac{dr}{\sqrt{1+k_y^2+k_z^2}},
\notag \\
\n
where $k_y=dy/dx$ and $k_z=dz/dx$.
Using Eq.~\eqref{R1} and the definition \eqref{la}, we obtain
\e
\label{laphi}
\begin{aligned}
\lambda(\phi)  &=(k_y^2+k_z^2+1)^{-1} A^{-4} \left((e+1)^2 \left(A^2 k_y^2+B^2\right)\right.
\\
&+\left(A^2 (k_z-\cot (\theta ))+B (e+1) \cot (\theta )\right)^2).
\end{aligned}
\q
which is independent of $a$, where we have used the short-hand notations $A=e \cos (\phi  \sin (\theta ))+1$ and $B=(e+1) \cos (\phi  \sin (\theta ))+e k_y\sin (\theta ) \sin (\phi  \sin (\theta ))$. The function $\lambda(\phi)$ is periodic with the same period $T_3$ as $R$, as shown in Fig.~\ref{fig:la}. The ratio of the distance to the initial one decreases and increases periodically for some ranges of $\phi$.  For $dl \neq 0$, or $k_z \neq 0$, two orbits $\bm{R}(\phi)$ and $\bm{R}_1(\phi)$ are confined to different cones. When $dl=0$ which corresponds to two orbits $\bm{R}(\phi)$ and $\bm{R}_1(\phi)$ confined to the same cone, $\lambda(\phi)$ is given by
\e
\label{laphi}
\begin{aligned}
\lambda(\phi)  &=\left(\cot ^2(\theta )+k_y^2+1\right)^{-1} A^{-4} (e+1)^2 \left(A^2 k_y^2+B^2 \csc ^2(\theta )\right).
\end{aligned}
\q
When $k_y=0,k_z=\cot (\theta)$, we have
\m
dl=d\phi=0,
\n
\m
\lambda(\phi)=\frac{(e+1)^4 \cos ^2(\phi  \sin (\theta ))}{(e \cos (\phi  \sin (\theta ))+1)^4},
\n
which means that two orbits are confined to the same cone, and when $R_1=R$, two orbits intersect, that is, $\lambda=0$. By analyzing Eq.~\eqref{laphi}, we conclude that only when $k_y=0$, $k_z=\cot (\theta)$, $\phi=\frac{(\frac{1}{2}+l)\pi}{\sin \theta}$, where $l$ is an integer, we have $\lambda=0$. For other cases, $\lambda>0$. By using Eqs.~\eqref{dphidt} and \eqref{laphi}, we plot $\lambda (\phi(t))$ as function of $t$ by choosing $ \theta=\pi/2\times 0.3, e=0.5$ in Fig.~\ref{fig:la}.  Dashed lines represent the case $dl=0$ while solid lines represent the case $dl \neq 0$.

For the case with $dx = 0$ and $dy \neq0$, we can set
\m
\bm{R_1}(0)-\bm{R}(0)=(dx, dy, dz)=(0,1, k_z)\frac{dr}{\sqrt{1+k_z^2}}.
\notag \\
\n
By using Eq.~\eqref{R1}, $\lambda$ can be expressed as
\m
&&\lambda (\phi)=(k_z^2+1)^{-1} A^{-4}
\notag \\
&\times& ((e+1)^2 \left(e^2 \sin ^2(\theta ) \sin ^2(\phi  \sin (\theta ))+A^2\right)
\notag \\
&+&(e (e+1) \cos (\theta ) \sin (\phi  \sin (\theta ))
+k_z A^2)^2).
\notag \\
\n

For the case with $dx =0$, $dy = 0$, $dz \neq 0$, we have
\m
 da=de=0, \quad dl=-dz,
\n
\m
\bm{R}_1(\phi)-\bm{R}(\phi)=(0,0,dz),
\n
which means that $\lambda(\phi)=1$ is a constant. We have observed, as expected for the conserved autonomous systems, that the relative ratio of two neighboring orbits during their evolutions is periodic as a function of $\phi$ and also as a function of time. For each period, there are exponential growths (decreases) in time.

\section*{Appendix B: The loss rate of $\tilde{L}$}
When we consider gravitational and electromagnetic radiations, we find only $\dot{L}^3 \neq 0$, which implies that the direction of $\bm{L}$ does not change and the magnitude of $\bm{L}$ decreases,  while the direction and magnitude of $\bm{\tilde{L}}$ both change. Here, we show how to use $d\tilde{L}/dt$ to get the same results of $da/dt$ and $d\theta/dt$.
From Eq.~\eqref{Relation}, we obtain
\m
\left\langle\frac{d\tilde{L}}{dt}\right\rangle=\left\langle\frac{dL}{dt}\right\rangle \frac{1}{\sin \theta}.
\n
From $\tilde{L}=\sqrt{-\mu Ca}$, we have the time revolution of the semimajor axis
\m
\label{dadt}
\frac{da}{dt}&=& \frac{4 C ((\Delta \sigma_q)^{2}+(\Delta \sigma_g)^{2}) \csc ^2(\theta )}{3 a^2}
\notag \\
&+& \frac{2 C^2 (15 \cos (2 \theta )-17) \csc ^4(\theta )}{5 a^3  \mu}.
\n
According to $\tilde{L}=\sqrt{-\mu Ca}$, $\tan (\theta)=\frac{\sqrt{-\mu C a}}{|D|}$, we have
\m
a=\frac{D^2}{-\mu C} \tan ^2(\theta),
\n
\m
\frac{da}{dt}=2 a \csc (\theta ) \sec (\theta ) \frac{d\theta}{dt}.
\n
Thus we can find the time evolution of the conic angle:
\m
\frac{d\theta}{dt}&=& \frac{2 C (\Delta \sigma_q)^{2}+(\Delta \sigma_g)^{2}) \cot (\theta )}{3 a^3}
\notag \\
&+& \frac{C^2 (15 \cos (2 \theta )-17) \cot (\theta ) \csc ^2(\theta )}{5 a^4 \mu }.
\n


\bibliography{merger_EB_20-07-19}

\begin{thebibliography}{43}%
\makeatletter
\providecommand \@ifxundefined [1]{%
 \@ifx{#1\undefined}
}%
\providecommand \@ifnum [1]{%
 \ifnum #1\expandafter \@firstoftwo
 \else \expandafter \@secondoftwo
 \fi
}%
\providecommand \@ifx [1]{%
 \ifx #1\expandafter \@firstoftwo
 \else \expandafter \@secondoftwo
 \fi
}%
\providecommand \natexlab [1]{#1}%
\providecommand \enquote  [1]{``#1''}%
\providecommand \bibnamefont  [1]{#1}%
\providecommand \bibfnamefont [1]{#1}%
\providecommand \citenamefont [1]{#1}%
\providecommand \href@noop [0]{\@secondoftwo}%
\providecommand \href [0]{\begingroup \@sanitize@url \@href}%
\providecommand \@href[1]{\@@startlink{#1}\@@href}%
\providecommand \@@href[1]{\endgroup#1\@@endlink}%
\providecommand \@sanitize@url [0]{\catcode `\\12\catcode `\$12\catcode
  `\&12\catcode `\#12\catcode `\^12\catcode `\_12\catcode `\%12\relax}%
\providecommand \@@startlink[1]{}%
\providecommand \@@endlink[0]{}%
\providecommand \url  [0]{\begingroup\@sanitize@url \@url }%
\providecommand \@url [1]{\endgroup\@href {#1}{\urlprefix }}%
\providecommand \urlprefix  [0]{URL }%
\providecommand \Eprint [0]{\href }%
\providecommand \doibase [0]{http://dx.doi.org/}%
\providecommand \selectlanguage [0]{\@gobble}%
\providecommand \bibinfo  [0]{\@secondoftwo}%
\providecommand \bibfield  [0]{\@secondoftwo}%
\providecommand \translation [1]{[#1]}%
\providecommand \BibitemOpen [0]{}%
\providecommand \bibitemStop [0]{}%
\providecommand \bibitemNoStop [0]{.\EOS\space}%
\providecommand \EOS [0]{\spacefactor3000\relax}%
\providecommand \BibitemShut  [1]{\csname bibitem#1\endcsname}%
\let\auto@bib@innerbib\@empty
\bibitem [{\citenamefont {Abbott}\ \emph
  {et~al.}(2016{\natexlab{a}})\citenamefont {Abbott} \emph
  {et~al.}}]{Abbott:2016blz}%
  \BibitemOpen
  \bibfield  {author} {\bibinfo {author} {\bibfnamefont {B.~P.}\ \bibnamefont
  {Abbott}} \emph {et~al.} (\bibinfo {collaboration} {LIGO Scientific,
  Virgo}),\ }\href {\doibase 10.1103/PhysRevLett.116.061102} {\bibfield
  {journal} {\bibinfo  {journal} {Phys. Rev. Lett.}\ }\textbf {\bibinfo
  {volume} {116}},\ \bibinfo {pages} {061102} (\bibinfo {year}
  {2016}{\natexlab{a}})},\ \Eprint {http://arxiv.org/abs/1602.03837}
  {arXiv:1602.03837 [gr-qc]} \BibitemShut {NoStop}%
\bibitem [{\citenamefont {Abbott}\ \emph
  {et~al.}(2016{\natexlab{b}})\citenamefont {Abbott} \emph
  {et~al.}}]{Abbott:2016nmj}%
  \BibitemOpen
  \bibfield  {author} {\bibinfo {author} {\bibfnamefont {B.~P.}\ \bibnamefont
  {Abbott}} \emph {et~al.} (\bibinfo {collaboration} {LIGO Scientific,
  Virgo}),\ }\href {\doibase 10.1103/PhysRevLett.116.241103} {\bibfield
  {journal} {\bibinfo  {journal} {Phys. Rev. Lett.}\ }\textbf {\bibinfo
  {volume} {116}},\ \bibinfo {pages} {241103} (\bibinfo {year}
  {2016}{\natexlab{b}})},\ \Eprint {http://arxiv.org/abs/1606.04855}
  {arXiv:1606.04855 [gr-qc]} \BibitemShut {NoStop}%
\bibitem [{\citenamefont {Abbott}\ \emph
  {et~al.}(2016{\natexlab{c}})\citenamefont {Abbott} \emph
  {et~al.}}]{TheLIGOScientific:2016pea}%
  \BibitemOpen
  \bibfield  {author} {\bibinfo {author} {\bibfnamefont {B.~P.}\ \bibnamefont
  {Abbott}} \emph {et~al.} (\bibinfo {collaboration} {LIGO Scientific,
  Virgo}),\ }\href {\doibase 10.1103/PhysRevX.6.041015,
  10.1103/PhysRevX.8.039903} {\bibfield  {journal} {\bibinfo  {journal} {Phys.
  Rev.}\ }\textbf {\bibinfo {volume} {X6}},\ \bibinfo {pages} {041015}
  (\bibinfo {year} {2016}{\natexlab{c}})},\ \bibinfo {note} {[erratum: Phys.
  Rev.X8,no.3,039903(2018)]},\ \Eprint {http://arxiv.org/abs/1606.04856}
  {arXiv:1606.04856 [gr-qc]} \BibitemShut {NoStop}%
\bibitem [{\citenamefont {Abbott}\ \emph
  {et~al.}(2017{\natexlab{a}})\citenamefont {Abbott} \emph
  {et~al.}}]{Abbott:2017vtc}%
  \BibitemOpen
  \bibfield  {author} {\bibinfo {author} {\bibfnamefont {B.~P.}\ \bibnamefont
  {Abbott}} \emph {et~al.} (\bibinfo {collaboration} {LIGO Scientific,
  VIRGO}),\ }\href {\doibase 10.1103/PhysRevLett.118.221101,
  10.1103/PhysRevLett.121.129901} {\bibfield  {journal} {\bibinfo  {journal}
  {Phys. Rev. Lett.}\ }\textbf {\bibinfo {volume} {118}},\ \bibinfo {pages}
  {221101} (\bibinfo {year} {2017}{\natexlab{a}})},\ \bibinfo {note} {[Erratum:
  Phys. Rev. Lett.121,no.12,129901(2018)]},\ \Eprint
  {http://arxiv.org/abs/1706.01812} {arXiv:1706.01812 [gr-qc]} \BibitemShut
  {NoStop}%
\bibitem [{\citenamefont {Abbott}\ \emph
  {et~al.}(2017{\natexlab{b}})\citenamefont {Abbott} \emph
  {et~al.}}]{Abbott:2017gyy}%
  \BibitemOpen
  \bibfield  {author} {\bibinfo {author} {\bibfnamefont {B.~P.}\ \bibnamefont
  {Abbott}} \emph {et~al.} (\bibinfo {collaboration} {LIGO Scientific,
  Virgo}),\ }\href {\doibase 10.3847/2041-8213/aa9f0c} {\bibfield  {journal}
  {\bibinfo  {journal} {Astrophys. J.}\ }\textbf {\bibinfo {volume} {851}},\
  \bibinfo {pages} {L35} (\bibinfo {year} {2017}{\natexlab{b}})},\ \Eprint
  {http://arxiv.org/abs/1711.05578} {arXiv:1711.05578 [astro-ph.HE]}
  \BibitemShut {NoStop}%
\bibitem [{\citenamefont {Abbott}\ \emph
  {et~al.}(2017{\natexlab{c}})\citenamefont {Abbott} \emph
  {et~al.}}]{Abbott:2017oio}%
  \BibitemOpen
  \bibfield  {author} {\bibinfo {author} {\bibfnamefont {B.~P.}\ \bibnamefont
  {Abbott}} \emph {et~al.} (\bibinfo {collaboration} {LIGO Scientific,
  Virgo}),\ }\href {\doibase 10.1103/PhysRevLett.119.141101} {\bibfield
  {journal} {\bibinfo  {journal} {Phys. Rev. Lett.}\ }\textbf {\bibinfo
  {volume} {119}},\ \bibinfo {pages} {141101} (\bibinfo {year}
  {2017}{\natexlab{c}})},\ \Eprint {http://arxiv.org/abs/1709.09660}
  {arXiv:1709.09660 [gr-qc]} \BibitemShut {NoStop}%
\bibitem [{\citenamefont {Abbott}\ \emph {et~al.}(2019)\citenamefont {Abbott}
  \emph {et~al.}}]{LIGOScientific:2018mvr}%
  \BibitemOpen
  \bibfield  {author} {\bibinfo {author} {\bibfnamefont {B.~P.}\ \bibnamefont
  {Abbott}} \emph {et~al.} (\bibinfo {collaboration} {LIGO Scientific,
  Virgo}),\ }\href {\doibase 10.1103/PhysRevX.9.031040} {\bibfield  {journal}
  {\bibinfo  {journal} {Phys. Rev.}\ }\textbf {\bibinfo {volume} {X9}},\
  \bibinfo {pages} {031040} (\bibinfo {year} {2019})},\ \Eprint
  {http://arxiv.org/abs/1811.12907} {arXiv:1811.12907 [astro-ph.HE]}
  \BibitemShut {NoStop}%
\bibitem [{\citenamefont {Bird}\ \emph {et~al.}(2016)\citenamefont {Bird},
  \citenamefont {Cholis}, \citenamefont {Muñoz}, \citenamefont {Ali-Haïmoud},
  \citenamefont {Kamionkowski}, \citenamefont {Kovetz}, \citenamefont
  {Raccanelli},\ and\ \citenamefont {Riess}}]{Bird:2016dcv}%
  \BibitemOpen
  \bibfield  {author} {\bibinfo {author} {\bibfnamefont {S.}~\bibnamefont
  {Bird}}, \bibinfo {author} {\bibfnamefont {I.}~\bibnamefont {Cholis}},
  \bibinfo {author} {\bibfnamefont {J.~B.}\ \bibnamefont {Muñoz}}, \bibinfo
  {author} {\bibfnamefont {Y.}~\bibnamefont {Ali-Haïmoud}}, \bibinfo {author}
  {\bibfnamefont {M.}~\bibnamefont {Kamionkowski}}, \bibinfo {author}
  {\bibfnamefont {E.~D.}\ \bibnamefont {Kovetz}}, \bibinfo {author}
  {\bibfnamefont {A.}~\bibnamefont {Raccanelli}}, \ and\ \bibinfo {author}
  {\bibfnamefont {A.~G.}\ \bibnamefont {Riess}},\ }\href {\doibase
  10.1103/PhysRevLett.116.201301} {\bibfield  {journal} {\bibinfo  {journal}
  {Phys. Rev. Lett.}\ }\textbf {\bibinfo {volume} {116}},\ \bibinfo {pages}
  {201301} (\bibinfo {year} {2016})},\ \Eprint
  {http://arxiv.org/abs/1603.00464} {arXiv:1603.00464 [astro-ph.CO]}
  \BibitemShut {NoStop}%
\bibitem [{\citenamefont {Sasaki}\ \emph {et~al.}(2016)\citenamefont {Sasaki},
  \citenamefont {Suyama}, \citenamefont {Tanaka},\ and\ \citenamefont
  {Yokoyama}}]{Sasaki:2016jop}%
  \BibitemOpen
  \bibfield  {author} {\bibinfo {author} {\bibfnamefont {M.}~\bibnamefont
  {Sasaki}}, \bibinfo {author} {\bibfnamefont {T.}~\bibnamefont {Suyama}},
  \bibinfo {author} {\bibfnamefont {T.}~\bibnamefont {Tanaka}}, \ and\ \bibinfo
  {author} {\bibfnamefont {S.}~\bibnamefont {Yokoyama}},\ }\href {\doibase
  10.1103/PhysRevLett.121.059901, 10.1103/PhysRevLett.117.061101} {\bibfield
  {journal} {\bibinfo  {journal} {Phys. Rev. Lett.}\ }\textbf {\bibinfo
  {volume} {117}},\ \bibinfo {pages} {061101} (\bibinfo {year} {2016})},\
  \bibinfo {note} {[erratum: Phys. Rev. Lett.121,no.5,059901(2018)]},\ \Eprint
  {http://arxiv.org/abs/1603.08338} {arXiv:1603.08338 [astro-ph.CO]}
  \BibitemShut {NoStop}%
\bibitem [{\citenamefont {Clesse}\ and\ \citenamefont
  {García-Bellido}(2017)}]{Clesse:2016vqa}%
  \BibitemOpen
  \bibfield  {author} {\bibinfo {author} {\bibfnamefont {S.}~\bibnamefont
  {Clesse}}\ and\ \bibinfo {author} {\bibfnamefont {J.}~\bibnamefont
  {García-Bellido}},\ }\href {\doibase 10.1016/j.dark.2016.10.002} {\bibfield
  {journal} {\bibinfo  {journal} {Phys. Dark Univ.}\ }\textbf {\bibinfo
  {volume} {15}},\ \bibinfo {pages} {142} (\bibinfo {year} {2017})},\ \Eprint
  {http://arxiv.org/abs/1603.05234} {arXiv:1603.05234 [astro-ph.CO]}
  \BibitemShut {NoStop}%
\bibitem [{\citenamefont {Ali-Haïmoud}\ \emph {et~al.}(2017)\citenamefont
  {Ali-Haïmoud}, \citenamefont {Kovetz},\ and\ \citenamefont
  {Kamionkowski}}]{Ali-Haimoud:2017rtz}%
  \BibitemOpen
  \bibfield  {author} {\bibinfo {author} {\bibfnamefont {Y.}~\bibnamefont
  {Ali-Haïmoud}}, \bibinfo {author} {\bibfnamefont {E.~D.}\ \bibnamefont
  {Kovetz}}, \ and\ \bibinfo {author} {\bibfnamefont {M.}~\bibnamefont
  {Kamionkowski}},\ }\href {\doibase 10.1103/PhysRevD.96.123523} {\bibfield
  {journal} {\bibinfo  {journal} {Phys. Rev.}\ }\textbf {\bibinfo {volume}
  {D96}},\ \bibinfo {pages} {123523} (\bibinfo {year} {2017})},\ \Eprint
  {http://arxiv.org/abs/1709.06576} {arXiv:1709.06576 [astro-ph.CO]}
  \BibitemShut {NoStop}%
\bibitem [{\citenamefont {Belczynski}\ \emph {et~al.}(2016)\citenamefont
  {Belczynski}, \citenamefont {Holz}, \citenamefont {Bulik},\ and\
  \citenamefont {O'Shaughnessy}}]{Belczynski:2016obo}%
  \BibitemOpen
  \bibfield  {author} {\bibinfo {author} {\bibfnamefont {K.}~\bibnamefont
  {Belczynski}}, \bibinfo {author} {\bibfnamefont {D.~E.}\ \bibnamefont
  {Holz}}, \bibinfo {author} {\bibfnamefont {T.}~\bibnamefont {Bulik}}, \ and\
  \bibinfo {author} {\bibfnamefont {R.}~\bibnamefont {O'Shaughnessy}},\ }\href
  {\doibase 10.1038/nature18322} {\bibfield  {journal} {\bibinfo  {journal}
  {Nature}\ }\textbf {\bibinfo {volume} {534}},\ \bibinfo {pages} {512}
  (\bibinfo {year} {2016})},\ \Eprint {http://arxiv.org/abs/1602.04531}
  {arXiv:1602.04531 [astro-ph.HE]} \BibitemShut {NoStop}%
\bibitem [{\citenamefont {Casares}\ and\ \citenamefont
  {Jonker}(2014)}]{Casares:2013tpa}%
  \BibitemOpen
  \bibfield  {author} {\bibinfo {author} {\bibfnamefont {J.}~\bibnamefont
  {Casares}}\ and\ \bibinfo {author} {\bibfnamefont {P.~G.}\ \bibnamefont
  {Jonker}},\ }\href {\doibase 10.1007/s11214-013-0030-6} {\bibfield  {journal}
  {\bibinfo  {journal} {Space Sci. Rev.}\ }\textbf {\bibinfo {volume} {183}},\
  \bibinfo {pages} {223} (\bibinfo {year} {2014})},\ \Eprint
  {http://arxiv.org/abs/1311.5118} {arXiv:1311.5118 [astro-ph.HE]} \BibitemShut
  {NoStop}%
\bibitem [{\citenamefont {Corral-Santana}\ \emph {et~al.}(2013)\citenamefont
  {Corral-Santana}, \citenamefont {Casares}, \citenamefont {Muñoz-Darias},
  \citenamefont {Rodríguez-Gil}, \citenamefont {Shahbaz}, \citenamefont
  {Torres}, \citenamefont {Zurita},\ and\ \citenamefont
  {Tyndall}}]{Corral-Santana:2013uua}%
  \BibitemOpen
  \bibfield  {author} {\bibinfo {author} {\bibfnamefont {J.~M.}\ \bibnamefont
  {Corral-Santana}}, \bibinfo {author} {\bibfnamefont {J.}~\bibnamefont
  {Casares}}, \bibinfo {author} {\bibfnamefont {T.}~\bibnamefont
  {Muñoz-Darias}}, \bibinfo {author} {\bibfnamefont {P.}~\bibnamefont
  {Rodríguez-Gil}}, \bibinfo {author} {\bibfnamefont {T.}~\bibnamefont
  {Shahbaz}}, \bibinfo {author} {\bibfnamefont {M.~A.~P.}\ \bibnamefont
  {Torres}}, \bibinfo {author} {\bibfnamefont {C.}~\bibnamefont {Zurita}}, \
  and\ \bibinfo {author} {\bibfnamefont {A.~A.}\ \bibnamefont {Tyndall}},\
  }\href {\doibase 10.1126/science.1228222} {\bibfield  {journal} {\bibinfo
  {journal} {Science}\ }\textbf {\bibinfo {volume} {339}},\ \bibinfo {pages}
  {1048} (\bibinfo {year} {2013})},\ \Eprint {http://arxiv.org/abs/1303.0034}
  {arXiv:1303.0034 [astro-ph.GA]} \BibitemShut {NoStop}%
\bibitem [{\citenamefont {Corral-Santana}\ \emph {et~al.}(2016)\citenamefont
  {Corral-Santana}, \citenamefont {Casares}, \citenamefont {Munoz-Darias},
  \citenamefont {Bauer}, \citenamefont {Martinez-Pais},\ and\ \citenamefont
  {Russell}}]{Corral-Santana:2015fud}%
  \BibitemOpen
  \bibfield  {author} {\bibinfo {author} {\bibfnamefont {J.~M.}\ \bibnamefont
  {Corral-Santana}}, \bibinfo {author} {\bibfnamefont {J.}~\bibnamefont
  {Casares}}, \bibinfo {author} {\bibfnamefont {T.}~\bibnamefont
  {Munoz-Darias}}, \bibinfo {author} {\bibfnamefont {F.~E.}\ \bibnamefont
  {Bauer}}, \bibinfo {author} {\bibfnamefont {I.~G.}\ \bibnamefont
  {Martinez-Pais}}, \ and\ \bibinfo {author} {\bibfnamefont {D.~M.}\
  \bibnamefont {Russell}},\ }\href {\doibase 10.1051/0004-6361/201527130}
  {\bibfield  {journal} {\bibinfo  {journal} {Astron. Astrophys.}\ }\textbf
  {\bibinfo {volume} {587}},\ \bibinfo {pages} {A61} (\bibinfo {year}
  {2016})},\ \Eprint {http://arxiv.org/abs/1510.08869} {arXiv:1510.08869
  [astro-ph.HE]} \BibitemShut {NoStop}%
\bibitem [{\citenamefont {Chen}\ and\ \citenamefont
  {Huang}(2018)}]{Chen:2018czv}%
  \BibitemOpen
  \bibfield  {author} {\bibinfo {author} {\bibfnamefont {Z.-C.}\ \bibnamefont
  {Chen}}\ and\ \bibinfo {author} {\bibfnamefont {Q.-G.}\ \bibnamefont
  {Huang}},\ }\href {\doibase 10.3847/1538-4357/aad6e2} {\bibfield  {journal}
  {\bibinfo  {journal} {Astrophys. J.}\ }\textbf {\bibinfo {volume} {864}},\
  \bibinfo {pages} {61} (\bibinfo {year} {2018})},\ \Eprint
  {http://arxiv.org/abs/1801.10327} {arXiv:1801.10327 [astro-ph.CO]}
  \BibitemShut {NoStop}%
\bibitem [{\citenamefont {Raidal}\ \emph {et~al.}(2019)\citenamefont {Raidal},
  \citenamefont {Spethmann}, \citenamefont {Vaskonen},\ and\ \citenamefont
  {Veermäe}}]{Raidal:2018bbj}%
  \BibitemOpen
  \bibfield  {author} {\bibinfo {author} {\bibfnamefont {M.}~\bibnamefont
  {Raidal}}, \bibinfo {author} {\bibfnamefont {C.}~\bibnamefont {Spethmann}},
  \bibinfo {author} {\bibfnamefont {V.}~\bibnamefont {Vaskonen}}, \ and\
  \bibinfo {author} {\bibfnamefont {H.}~\bibnamefont {Veermäe}},\ }\href
  {\doibase 10.1088/1475-7516/2019/02/018} {\bibfield  {journal} {\bibinfo
  {journal} {JCAP}\ }\textbf {\bibinfo {volume} {1902}},\ \bibinfo {pages}
  {018} (\bibinfo {year} {2019})},\ \Eprint {http://arxiv.org/abs/1812.01930}
  {arXiv:1812.01930 [astro-ph.CO]} \BibitemShut {NoStop}%
\bibitem [{\citenamefont {Liu}\ \emph {et~al.}(2019{\natexlab{a}})\citenamefont
  {Liu}, \citenamefont {Guo},\ and\ \citenamefont {Cai}}]{Liu:2018ess}%
  \BibitemOpen
  \bibfield  {author} {\bibinfo {author} {\bibfnamefont {L.}~\bibnamefont
  {Liu}}, \bibinfo {author} {\bibfnamefont {Z.-K.}\ \bibnamefont {Guo}}, \ and\
  \bibinfo {author} {\bibfnamefont {R.-G.}\ \bibnamefont {Cai}},\ }\href
  {\doibase 10.1103/PhysRevD.99.063523} {\bibfield  {journal} {\bibinfo
  {journal} {Phys. Rev.}\ }\textbf {\bibinfo {volume} {D99}},\ \bibinfo {pages}
  {063523} (\bibinfo {year} {2019}{\natexlab{a}})},\ \Eprint
  {http://arxiv.org/abs/1812.05376} {arXiv:1812.05376 [astro-ph.CO]}
  \BibitemShut {NoStop}%
\bibitem [{\citenamefont {Liu}\ \emph {et~al.}(2019{\natexlab{b}})\citenamefont
  {Liu}, \citenamefont {Guo},\ and\ \citenamefont {Cai}}]{Liu:2019rnx}%
  \BibitemOpen
  \bibfield  {author} {\bibinfo {author} {\bibfnamefont {L.}~\bibnamefont
  {Liu}}, \bibinfo {author} {\bibfnamefont {Z.-K.}\ \bibnamefont {Guo}}, \ and\
  \bibinfo {author} {\bibfnamefont {R.-G.}\ \bibnamefont {Cai}},\ }\href
  {\doibase 10.1140/epjc/s10052-019-7227-0} {\bibfield  {journal} {\bibinfo
  {journal} {Eur. Phys. J.}\ }\textbf {\bibinfo {volume} {C79}},\ \bibinfo
  {pages} {717} (\bibinfo {year} {2019}{\natexlab{b}})},\ \Eprint
  {http://arxiv.org/abs/1901.07672} {arXiv:1901.07672 [astro-ph.CO]}
  \BibitemShut {NoStop}%
\bibitem [{\citenamefont {Hawking}(1971)}]{Hawking:1971ei}%
  \BibitemOpen
  \bibfield  {author} {\bibinfo {author} {\bibfnamefont {S.}~\bibnamefont
  {Hawking}},\ }\href@noop {} {\bibfield  {journal} {\bibinfo  {journal} {Mon.
  Not. Roy. Astron. Soc.}\ }\textbf {\bibinfo {volume} {152}},\ \bibinfo
  {pages} {75} (\bibinfo {year} {1971})}\BibitemShut {NoStop}%
\bibitem [{\citenamefont {Carr}\ and\ \citenamefont
  {Hawking}(1974)}]{Carr:1974nx}%
  \BibitemOpen
  \bibfield  {author} {\bibinfo {author} {\bibfnamefont {B.~J.}\ \bibnamefont
  {Carr}}\ and\ \bibinfo {author} {\bibfnamefont {S.~W.}\ \bibnamefont
  {Hawking}},\ }\href@noop {} {\bibfield  {journal} {\bibinfo  {journal} {Mon.
  Not. Roy. Astron. Soc.}\ }\textbf {\bibinfo {volume} {168}},\ \bibinfo
  {pages} {399} (\bibinfo {year} {1974})}\BibitemShut {NoStop}%
\bibitem [{\citenamefont {Bean}\ and\ \citenamefont
  {Magueijo}(2002)}]{Bean:2002kx}%
  \BibitemOpen
  \bibfield  {author} {\bibinfo {author} {\bibfnamefont {R.}~\bibnamefont
  {Bean}}\ and\ \bibinfo {author} {\bibfnamefont {J.}~\bibnamefont
  {Magueijo}},\ }\href {\doibase 10.1103/PhysRevD.66.063505} {\bibfield
  {journal} {\bibinfo  {journal} {Phys. Rev.}\ }\textbf {\bibinfo {volume}
  {D66}},\ \bibinfo {pages} {063505} (\bibinfo {year} {2002})},\ \Eprint
  {http://arxiv.org/abs/astro-ph/0204486} {arXiv:astro-ph/0204486 [astro-ph]}
  \BibitemShut {NoStop}%
\bibitem [{\citenamefont {Kawasaki}\ \emph {et~al.}(2012)\citenamefont
  {Kawasaki}, \citenamefont {Kusenko},\ and\ \citenamefont
  {Yanagida}}]{Kawasaki:2012kn}%
  \BibitemOpen
  \bibfield  {author} {\bibinfo {author} {\bibfnamefont {M.}~\bibnamefont
  {Kawasaki}}, \bibinfo {author} {\bibfnamefont {A.}~\bibnamefont {Kusenko}}, \
  and\ \bibinfo {author} {\bibfnamefont {T.~T.}\ \bibnamefont {Yanagida}},\
  }\href {\doibase 10.1016/j.physletb.2012.03.056} {\bibfield  {journal}
  {\bibinfo  {journal} {Phys. Lett.}\ }\textbf {\bibinfo {volume} {B711}},\
  \bibinfo {pages} {1} (\bibinfo {year} {2012})},\ \Eprint
  {http://arxiv.org/abs/1202.3848} {arXiv:1202.3848 [astro-ph.CO]} \BibitemShut
  {NoStop}%
\bibitem [{\citenamefont {Carr}\ and\ \citenamefont
  {Silk}(2018)}]{Carr:2018rid}%
  \BibitemOpen
  \bibfield  {author} {\bibinfo {author} {\bibfnamefont {B.}~\bibnamefont
  {Carr}}\ and\ \bibinfo {author} {\bibfnamefont {J.}~\bibnamefont {Silk}},\
  }\href {\doibase 10.1093/mnras/sty1204} {\bibfield  {journal} {\bibinfo
  {journal} {Mon. Not. Roy. Astron. Soc.}\ }\textbf {\bibinfo {volume} {478}},\
  \bibinfo {pages} {3756} (\bibinfo {year} {2018})},\ \Eprint
  {http://arxiv.org/abs/1801.00672} {arXiv:1801.00672 [astro-ph.CO]}
  \BibitemShut {NoStop}%
\bibitem [{\citenamefont {Cardoso}\ \emph {et~al.}(2016)\citenamefont
  {Cardoso}, \citenamefont {Macedo}, \citenamefont {Pani},\ and\ \citenamefont
  {Ferrari}}]{Cardoso:2016olt}%
  \BibitemOpen
  \bibfield  {author} {\bibinfo {author} {\bibfnamefont {V.}~\bibnamefont
  {Cardoso}}, \bibinfo {author} {\bibfnamefont {C.~F.~B.}\ \bibnamefont
  {Macedo}}, \bibinfo {author} {\bibfnamefont {P.}~\bibnamefont {Pani}}, \ and\
  \bibinfo {author} {\bibfnamefont {V.}~\bibnamefont {Ferrari}},\ }\href
  {\doibase 10.1088/1475-7516/2016/05/054, 10.1088/1475-7516/2020/04/E01}
  {\bibfield  {journal} {\bibinfo  {journal} {JCAP}\ }\textbf {\bibinfo
  {volume} {1605}},\ \bibinfo {pages} {054} (\bibinfo {year} {2016})},\
  \bibinfo {note} {[Erratum: JCAP2004,E01(2020)]},\ \Eprint
  {http://arxiv.org/abs/1604.07845} {arXiv:1604.07845 [hep-ph]} \BibitemShut
  {NoStop}%
\bibitem [{\citenamefont {Liebling}\ and\ \citenamefont
  {Palenzuela}(2016)}]{Liebling:2016orx}%
  \BibitemOpen
  \bibfield  {author} {\bibinfo {author} {\bibfnamefont {S.~L.}\ \bibnamefont
  {Liebling}}\ and\ \bibinfo {author} {\bibfnamefont {C.}~\bibnamefont
  {Palenzuela}},\ }\href {\doibase 10.1103/PhysRevD.94.064046} {\bibfield
  {journal} {\bibinfo  {journal} {Phys. Rev.}\ }\textbf {\bibinfo {volume}
  {D94}},\ \bibinfo {pages} {064046} (\bibinfo {year} {2016})},\ \Eprint
  {http://arxiv.org/abs/1607.02140} {arXiv:1607.02140 [gr-qc]} \BibitemShut
  {NoStop}%
\bibitem [{\citenamefont {Bai}\ and\ \citenamefont
  {Orlofsky}(2020)}]{Bai:2019zcd}%
  \BibitemOpen
  \bibfield  {author} {\bibinfo {author} {\bibfnamefont {Y.}~\bibnamefont
  {Bai}}\ and\ \bibinfo {author} {\bibfnamefont {N.}~\bibnamefont {Orlofsky}},\
  }\href {\doibase 10.1103/PhysRevD.101.055006} {\bibfield  {journal} {\bibinfo
   {journal} {Phys. Rev. D}\ }\textbf {\bibinfo {volume} {101}},\ \bibinfo
  {pages} {055006} (\bibinfo {year} {2020})},\ \Eprint
  {http://arxiv.org/abs/1906.04858} {arXiv:1906.04858 [hep-ph]} \BibitemShut
  {NoStop}%
\bibitem [{\citenamefont {Liu}\ \emph {et~al.}(2020)\citenamefont {Liu},
  \citenamefont {Guo}, \citenamefont {Cai},\ and\ \citenamefont
  {Kim}}]{Liu:2020cds}%
  \BibitemOpen
  \bibfield  {author} {\bibinfo {author} {\bibfnamefont {L.}~\bibnamefont
  {Liu}}, \bibinfo {author} {\bibfnamefont {Z.-K.}\ \bibnamefont {Guo}},
  \bibinfo {author} {\bibfnamefont {R.-G.}\ \bibnamefont {Cai}}, \ and\
  \bibinfo {author} {\bibfnamefont {S.~P.}\ \bibnamefont {Kim}},\ }\href
  {\doibase 10.1103/PhysRevD.102.043508} {\bibfield  {journal} {\bibinfo
  {journal} {Phys. Rev. D}\ }\textbf {\bibinfo {volume} {102}},\ \bibinfo
  {pages} {043508} (\bibinfo {year} {2020})},\ \Eprint
  {http://arxiv.org/abs/2001.02984} {arXiv:2001.02984 [astro-ph.CO]}
  \BibitemShut {NoStop}%
\bibitem [{\citenamefont {Christiansen}\ \emph {et~al.}(2020)\citenamefont
  {Christiansen}, \citenamefont {Jiménez},\ and\ \citenamefont
  {Mota}}]{Christiansen:2020pnv}%
  \BibitemOpen
  \bibfield  {author} {\bibinfo {author} {\bibfnamefont {{\O}.}~\bibnamefont
  {Christiansen}}, \bibinfo {author} {\bibfnamefont {J.~B.}\ \bibnamefont
  {Jiménez}}, \ and\ \bibinfo {author} {\bibfnamefont {D.~F.}\ \bibnamefont
  {Mota}},\ }\href@noop {} {\  (\bibinfo {year} {2020})},\ \Eprint
  {http://arxiv.org/abs/2003.11452} {arXiv:2003.11452 [gr-qc]} \BibitemShut
  {NoStop}%
\bibitem [{\citenamefont {Wang}\ \emph {et~al.}(2020)\citenamefont {Wang},
  \citenamefont {Li}, \citenamefont {Jiang}, \citenamefont {Hu},\ and\
  \citenamefont {Fan}}]{Wang:2020fra}%
  \BibitemOpen
  \bibfield  {author} {\bibinfo {author} {\bibfnamefont {H.-T.}\ \bibnamefont
  {Wang}}, \bibinfo {author} {\bibfnamefont {P.-C.}\ \bibnamefont {Li}},
  \bibinfo {author} {\bibfnamefont {J.-L.}\ \bibnamefont {Jiang}}, \bibinfo
  {author} {\bibfnamefont {Y.-M.}\ \bibnamefont {Hu}}, \ and\ \bibinfo {author}
  {\bibfnamefont {Y.-Z.}\ \bibnamefont {Fan}},\ }\href@noop {} {\  (\bibinfo
  {year} {2020})},\ \Eprint {http://arxiv.org/abs/2004.12421} {arXiv:2004.12421
  [gr-qc]} \BibitemShut {NoStop}%
\bibitem [{\citenamefont {Bozzola}\ and\ \citenamefont
  {Paschalidis}(2020)}]{Bozzola:2020mjx}%
  \BibitemOpen
  \bibfield  {author} {\bibinfo {author} {\bibfnamefont {G.}~\bibnamefont
  {Bozzola}}\ and\ \bibinfo {author} {\bibfnamefont {V.}~\bibnamefont
  {Paschalidis}},\ }\href@noop {} {\  (\bibinfo {year} {2020})},\ \Eprint
  {http://arxiv.org/abs/2006.15764} {arXiv:2006.15764 [gr-qc]} \BibitemShut
  {NoStop}%
\bibitem [{\citenamefont {Maldacena}(2020)}]{Maldacena:2020skw}%
  \BibitemOpen
  \bibfield  {author} {\bibinfo {author} {\bibfnamefont {J.}~\bibnamefont
  {Maldacena}},\ }\href@noop {} {\  (\bibinfo {year} {2020})},\ \Eprint
  {http://arxiv.org/abs/2004.06084} {arXiv:2004.06084 [hep-th]} \BibitemShut
  {NoStop}%
\bibitem [{\citenamefont {Bai}\ \emph {et~al.}(2020)\citenamefont {Bai},
  \citenamefont {Berger}, \citenamefont {Korwar},\ and\ \citenamefont
  {Orlofsky}}]{Bai:2020spd}%
  \BibitemOpen
  \bibfield  {author} {\bibinfo {author} {\bibfnamefont {Y.}~\bibnamefont
  {Bai}}, \bibinfo {author} {\bibfnamefont {J.}~\bibnamefont {Berger}},
  \bibinfo {author} {\bibfnamefont {M.}~\bibnamefont {Korwar}}, \ and\ \bibinfo
  {author} {\bibfnamefont {N.}~\bibnamefont {Orlofsky}},\ }\href@noop {} {\
  (\bibinfo {year} {2020})},\ \Eprint {http://arxiv.org/abs/2007.03703}
  {arXiv:2007.03703 [hep-ph]} \BibitemShut {NoStop}%
\bibitem [{\citenamefont {{Kasuya}}(1982)}]{1982PhRvD..25..995K}%
  \BibitemOpen
  \bibfield  {author} {\bibinfo {author} {\bibfnamefont {M.}~\bibnamefont
  {{Kasuya}}},\ }\href {\doibase 10.1103/PhysRevD.25.995} {\bibfield  {journal}
  {\bibinfo  {journal} {\prd}\ }\textbf {\bibinfo {volume} {25}},\ \bibinfo
  {pages} {995} (\bibinfo {year} {1982})}\BibitemShut {NoStop}%
\bibitem [{\citenamefont {{Leach}}\ and\ \citenamefont
  {{Flessas}}(2004)}]{2004math.ph...3028L}%
  \BibitemOpen
  \bibfield  {author} {\bibinfo {author} {\bibfnamefont {P.~G.~L.}\
  \bibnamefont {{Leach}}}\ and\ \bibinfo {author} {\bibfnamefont {G.~P.}\
  \bibnamefont {{Flessas}}},\ }\href@noop {} {\bibfield  {journal} {\bibinfo
  {journal} {arXiv e-prints}\ ,\ \bibinfo {eid} {math-ph/0403028}} (\bibinfo
  {year} {2004})},\ \Eprint {http://arxiv.org/abs/math-ph/0403028}
  {arXiv:math-ph/0403028 [math-ph]} \BibitemShut {NoStop}%
\bibitem [{\citenamefont {Poincar{\'e}}(1896)}]{poincare1896remarques}%
  \BibitemOpen
  \bibfield  {author} {\bibinfo {author} {\bibfnamefont {H.}~\bibnamefont
  {Poincar{\'e}}},\ }\href@noop {} {\bibfield  {journal} {\bibinfo  {journal}
  {Comptes Rendus Acad. Sci}\ }\textbf {\bibinfo {volume} {123}},\ \bibinfo
  {pages} {530} (\bibinfo {year} {1896})}\BibitemShut {NoStop}%
\bibitem [{\citenamefont {Amaro-Seoane}\ \emph {et~al.}(2017)\citenamefont
  {Amaro-Seoane} \emph {et~al.}}]{Audley:2017drz}%
  \BibitemOpen
  \bibfield  {author} {\bibinfo {author} {\bibfnamefont {P.}~\bibnamefont
  {Amaro-Seoane}} \emph {et~al.} (\bibinfo {collaboration} {LISA}),\
  }\href@noop {} {\  (\bibinfo {year} {2017})},\ \Eprint
  {http://arxiv.org/abs/1702.00786} {arXiv:1702.00786 [astro-ph.IM]}
  \BibitemShut {NoStop}%
\bibitem [{\citenamefont {Ruan}\ \emph {et~al.}(2020)\citenamefont {Ruan},
  \citenamefont {Guo}, \citenamefont {Cai},\ and\ \citenamefont
  {Zhang}}]{Guo:2018npi}%
  \BibitemOpen
  \bibfield  {author} {\bibinfo {author} {\bibfnamefont {W.-H.}\ \bibnamefont
  {Ruan}}, \bibinfo {author} {\bibfnamefont {Z.-K.}\ \bibnamefont {Guo}},
  \bibinfo {author} {\bibfnamefont {R.-G.}\ \bibnamefont {Cai}}, \ and\
  \bibinfo {author} {\bibfnamefont {Y.-Z.}\ \bibnamefont {Zhang}},\ }\href
  {\doibase 10.1142/S0217751X2050075X} {\bibfield  {journal} {\bibinfo
  {journal} {Int. J. Mod. Phys.}\ }\textbf {\bibinfo {volume} {A35}},\ \bibinfo
  {pages} {2050075} (\bibinfo {year} {2020})},\ \Eprint
  {http://arxiv.org/abs/1807.09495} {arXiv:1807.09495 [gr-qc]} \BibitemShut
  {NoStop}%
\bibitem [{\citenamefont {{Schwinger}}\ \emph {et~al.}(1976)\citenamefont
  {{Schwinger}}, \citenamefont {{Milton}}, \citenamefont {{Tsai}},
  \citenamefont {{Deraad}},\ and\ \citenamefont
  {{Clark}}}]{1976AnPhy.101..451S}%
  \BibitemOpen
  \bibfield  {author} {\bibinfo {author} {\bibfnamefont {J.}~\bibnamefont
  {{Schwinger}}}, \bibinfo {author} {\bibfnamefont {K.~A.}\ \bibnamefont
  {{Milton}}}, \bibinfo {author} {\bibfnamefont {W.-Y.}\ \bibnamefont
  {{Tsai}}}, \bibinfo {author} {\bibfnamefont {J.}~\bibnamefont {{Deraad}},
  \bibfnamefont {Lester~L.}}, \ and\ \bibinfo {author} {\bibfnamefont {D.~C.}\
  \bibnamefont {{Clark}}},\ }\href {\doibase 10.1016/0003-4916(76)90020-8}
  {\bibfield  {journal} {\bibinfo  {journal} {Annals of Physics}\ }\textbf
  {\bibinfo {volume} {101}},\ \bibinfo {pages} {451} (\bibinfo {year}
  {1976})}\BibitemShut {NoStop}%
\bibitem [{\citenamefont {{Zwanziger}}(1968)}]{1968PhRv..176.1480Z}%
  \BibitemOpen
  \bibfield  {author} {\bibinfo {author} {\bibfnamefont {D.}~\bibnamefont
  {{Zwanziger}}},\ }\href {\doibase 10.1103/PhysRev.176.1480} {\bibfield
  {journal} {\bibinfo  {journal} {Physical Review}\ }\textbf {\bibinfo {volume}
  {176}},\ \bibinfo {pages} {1480} (\bibinfo {year} {1968})}\BibitemShut
  {NoStop}%
\bibitem [{\citenamefont {Argyris}\ \emph {et~al.}(2015)\citenamefont
  {Argyris}, \citenamefont {Faust}, \citenamefont {Haase},\ and\ \citenamefont
  {Friedrich}}]{Argyris:2015311}%
  \BibitemOpen
  \bibfield  {author} {\bibinfo {author} {\bibfnamefont {J.~H.}\ \bibnamefont
  {Argyris}}, \bibinfo {author} {\bibfnamefont {G.}~\bibnamefont {Faust}},
  \bibinfo {author} {\bibfnamefont {M.}~\bibnamefont {Haase}}, \ and\ \bibinfo
  {author} {\bibfnamefont {R.}~\bibnamefont {Friedrich}},\ }\href {\doibase
  10.1007/978-3-662-46042-9} {\emph {\bibinfo {title} {{An exploration of
  dynamical systems and chaos; 2nd ed.}}}}\ (\bibinfo  {publisher} {Springer},\
  \bibinfo {address} {Berlin},\ \bibinfo {year} {2015})\BibitemShut {NoStop}%
\bibitem [{\citenamefont {Peters}\ and\ \citenamefont
  {Mathews}(1963)}]{Peters:1963ux}%
  \BibitemOpen
  \bibfield  {author} {\bibinfo {author} {\bibfnamefont {P.~C.}\ \bibnamefont
  {Peters}}\ and\ \bibinfo {author} {\bibfnamefont {J.}~\bibnamefont
  {Mathews}},\ }\href {\doibase 10.1103/PhysRev.131.435} {\bibfield  {journal}
  {\bibinfo  {journal} {Phys. Rev.}\ }\textbf {\bibinfo {volume} {131}},\
  \bibinfo {pages} {435} (\bibinfo {year} {1963})}\BibitemShut {NoStop}%
\bibitem [{\citenamefont {Peters}(1964)}]{Peters:1964zz}%
  \BibitemOpen
  \bibfield  {author} {\bibinfo {author} {\bibfnamefont {P.~C.}\ \bibnamefont
  {Peters}},\ }\href {\doibase 10.1103/PhysRev.136.B1224} {\bibfield  {journal}
  {\bibinfo  {journal} {Phys. Rev.}\ }\textbf {\bibinfo {volume} {136}},\
  \bibinfo {pages} {B1224} (\bibinfo {year} {1964})}\BibitemShut {NoStop}%
\end{thebibliography}%

\end{document}